%% file: main.tex
\documentclass[hidelinks,onefignum,onetabnum]{siamart250106}
\input{ex_shared}

\ifpdf
\hypersetup{
  pdftitle={Tensor train representations of Greeks for Fourier-based pricing of multi-asset options},
  pdfauthor={}
}
\fi
\usepackage{enumitem} 
\usepackage{color}
\usepackage{multirow}
\usepackage{booktabs}
\usepackage{dcolumn}  
\usepackage{bm}  
\usepackage[whole]{bxcjkjatype}   
\usepackage{verbatim}  
\usepackage{color} 
\usepackage{physics}   
\usepackage{subcaption}
\usepackage{graphicx,epstopdf} %

\usepackage{comment}
\usepackage{rotating}

\begin{document}
\maketitle
\setcounter{tocdepth}{2} 
\tableofcontents
\begin{abstract}
Efficient computation of Greeks for multi-asset options remains a key challenge in quantitative finance. 
While Monte Carlo (MC) simulation is widely used, 
it suffers from the large sample complexity for high accuracy.
We propose a framework to compute Greeks in a single evaluation of a tensor train (TT), which is obtained by compressing the Fourier transform (FT)-based pricing function via TT learning using tensor cross interpolation. 
Based on this TT representation, we introduce two approaches to compute Greeks: a numerical differentiation (ND) approach that applies a numerical differential operator to one tensor core and an analytical (AN) approach that constructs the TT of closed-form differentiation expressions of FT-based pricing. 
Numerical experiments on a five-asset min-call option in the Black-Sholes model show significant speed-ups of up to about $10^{5} \times$ over MC while maintaining comparable accuracy. 
The ND approach matches or exceeds the accuracy of the AN approach and requires lower computational complexity for constructing the TT representation, making it the preferred choice.
\end{abstract}



\section{Introduction}\label{sec:intro}
Pricing options has long been a central topic in quantitative finance.
Options are financial contracts whose payoffs are determined by prices of underlying assets such as stocks and bonds.
In addition to the option price itself, its derivatives with respect to the input parameters, such as underlying asset prices and volatilities, collectively known as sensitivities or option Greeks, are indispensable quantities for risk management.
Computing them accurately and efficiently is often challenging, especially for options on multiple assets, for which analytic formulas of prices and Greeks are usually unavailable. 
The Monte Carlo (MC) simulation, a common approach for multi-asset option pricing, requires generating many sample paths of underlying asset prices for accuracy, which leads to a long computational time.
This computational cost increases when computing Greeks if we simply use the finite difference (FD) method, in which we run multiple MC simulations with slightly different input parameters.
Although we can resort to the Malliavin calculus (MV)-based method\footnote{For a comprehensive review, readers are refered to Ref.~\cite{alos2021malliavin}.}, in which each of the Greeks is written as a single expectation, and we run a single MC simulation for it, the large sample complexity required for high accuracy remains inevitable.

Fourier-transform (FT)-based option pricing, which is the focus of this study, has been studied for many years as an alternative approach to option pricing~\cite{carr1999option, lewis2001simple}, particularly when the underlying stochastic process admits a analytical characteristic function (e.g., in the Black–Scholes (BS)~\cite{merton1973theory} or L\'{e}vy models~\cite{Applebaum_2001}). 
Given the characteristic function and the Fourier-transformed payoff function, this method gives an option price as an integral of their product.
Methods for computing Greeks by FT-based pricing have also been investigated~\cite{Schmelzle2010OptionPF, DeOliveira2016, Tobias2025}.
However, extending these methods to multi-asset option pricing combined with naive grid-based integration is challenging because it suffers from the curse of dimensionality, i.e., the exponential increase of computational time with respect to the asset number. 

Recently, tensor network (TN) representations have gained attention in physics and applied mathematics for compressing high-dimensional data. 
Among TN representations, the tensor train (TT)~\cite{Oseledets2011, ORUS2014117}, also known as the matrix product states~\cite{cmp/1104249404, Klumper1993, Schollwock2011-eq}, often proves effective at capturing low-dimensional structures in problems like fluid dynamics~\cite{Peddinti_2024, kornev2023numerical}, machine learning~\cite{novikov2016exponential, Stoudenmire2016, sakaue2025adaptivesamplingbasedoptimizationquantics}, chemical master equations~\cite{10.1063/5.0045521, Kinjo2025}, and quantum field theory~\cite{Nunez_Fernandez2022-fo, Shinaoka2023-lf,10.21468/SciPostPhys.18.1.007}, enabling faster operations such as integration, convolution, and Fourier transformation. 
Also, TN learning algorithms, such as tensor cross interpolation (TCI)~\cite{OSELEDETS201070, N_ez_Fern_ndez_2025}, have emerged for constructing an accurate TN representation of a target tensor from only a small subset of its entries, akin to machine learning.

FT-based pricing of multi-asset options also benefits from TT representation aided by TCI, as compressing high-dimensional characteristic functions and Fourier-transformed payoff functions in TT mitigates the curse of dimensionality~\cite{kastoryano2022highlyefficienttensornetwork}. 
One issue of this original TCI-aided FT-based method is that it is required to rerun TCI when we change the input parameters.
To address this issue, a recent study \cite{math13111828} by some of the authors incorporates the dependencies on input parameters directly into the TT representations of the functions used in FT-based pricing, allowing us to create a new function that takes the input parameters and outputs the option price.
This enables the rapid recalculation of option prices for varying parameters, offering significant speed-ups of multi-asset option pricing. 

In this study, we extend this previous work~\cite{math13111828}, which focused on computing option prices, to enable efficient evaluation of Greeks of multi-asset options.
As in the previous study, the proposed algorithm is decomposed into offline and online phases.
In the offline phase, we construct TT representations of FT-based Greeks with parameter dependencies incorporated.
In the online phase, Greeks are rapidly computed for varying parameters via TT contraction.

We propose two TT‑based methods for Greek calculations:
(1) a numerical differentiation (ND) approach that first constructs the TT representation of option price and then computes Greeks by directly applying numerical differentiation operators to its cores;
(2) an analytical differentiation (AN) approach that constructs separate TT representations explicitly for each of the Greeks from analytical expressions of FT-based option pricing. 
While the AN approach eliminates numerical differentiation errors, it necessitates constructing separate TT representations for each of the Greeks individually.

We demonstrate both approaches through pricing a five-asset min-call option under the Black–Scholes (BS) model with different correlation settings dubbed constant, constant with noise, and random.
The numerical results show that focusing on the online phase, the two proposed TT-based methods significantly outperform MC in terms of computational complexity and time, while maintaining comparable accuracy.
Moreover, the ND approach has lower computational complexity in the offline phase and achieves accuracy comparable to or better than that of the AN approach.

The rest of the article is organized as follows. Section~\ref{sec:TTandTTlearning} introduces the basics of the tensor-train (TT) representations and the TT learning algorithm. 
Section~\ref{sec:FT-basedpricing} provides an overview of FT-based option pricing and Greeks computations. 
In Section~\ref{GreeksTT}, we present the new methods.
Section~\ref{numericaldetails} presents the detailed setting of our numerical experiments, 
and Section~\ref{sec:experiments} reports the result. 
Finally, Section~\ref{sec:discussion} concludes the article, followed by listing future directions.

\section{Tensor train and learning algorithm}\label{sec:TTandTTlearning}
In this section, we introduce the tensor train and two compression techniques, TCI and singular value decomposition (SVD), used in this study. 

\subsection{Tensor train}
\begin{figure*}[htbp]
    \centering    \includegraphics[width=0.8\linewidth]{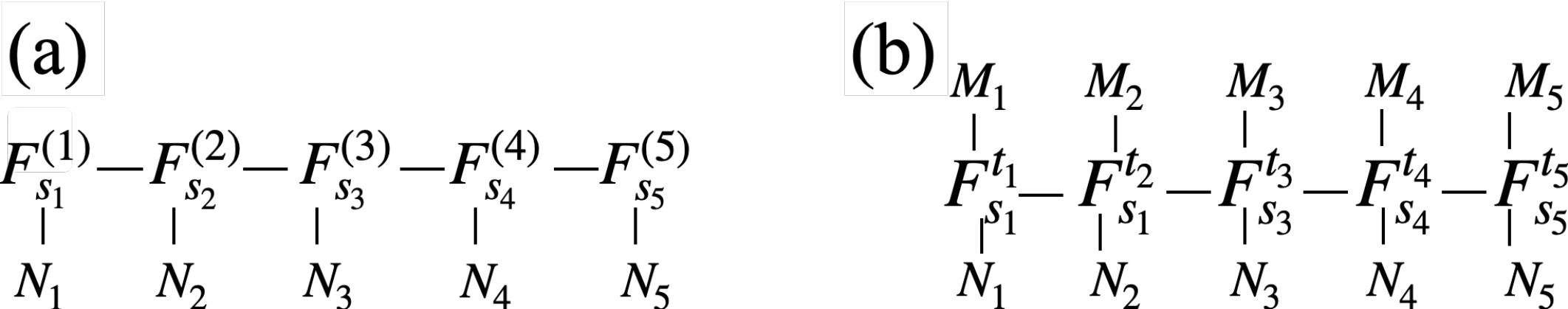}
    \caption{
    Diagram of (a) a tensor train and (b) a tensor train operator. 
    For clarity, at each site $i$ we show only the local index $s_i$ with its dimension $N_i$ (and for the (b), local index $t_i$ with its dimension $M_i$), omitting the bond indices and bond dimensions of the links between the cores.
    }
    \label{fig:tt_zu}
\end{figure*}

A $d$-way tensor $F_{s_1, \dots, s_d} \in \mathbb{R}^{N_1 \times N_2 \times \cdots \times N_d}$, where each local index $s_l$ has a local dimension $N_l$ ($l=1, \dots, d$), can be represented using a low-rank tensor decomposition. 
Assuming that correlations between distant variables are sufficiently weak while correlations between nearby variables are strong, tensor train, a kind of tensor network, can be applied to exploit this structure as follows. 
\begin{align}
    F_{s_1, \ldots, s_d} \approx 
     \tilde{F}_{s_1, \ldots, s_d}
     &=
    \sum_{a_1=1}^{\chi_1} \cdots \sum_{a_{d-1}=1}^{\chi_{d-1}} F_{a_0 s_1 a_1}^{(1)} F_{a_1 s_2 a_2}^{(2)} \cdots F_{a_{d-1} s_d a_d}^{(d)} 
    \equiv 
    \prod^{d}_{i} F_{s_i}^{(i)},
\end{align}
where $a_0$ and $a_d$ are called dummy indices whose bond dimension $\chi_0$ and $\chi_d$ are one, respectively. 
$F_{a_{i-1} s_i a_i}^{(i)}$ denotes each three-way tensor, often reffered to as a tensor core, $F_{a_{i-1}s_i a_{i}}^{(i)} \in \mathbb{R}^{\chi_{i-1} \times N_i \times \chi_i}$, $a_i$ represents the virtual bond index, and $\chi_i$ is the dimension of the virtual bond. 
Figure~\ref{fig:tt_zu} (a) shows the corresponding diagram.
One of the key advantages of TT is that it significantly reduces computational complexity and memory requirements by limiting the bond dimensions  $\chi_i$ to an upper bound $\chi$, provided that $F_{s_1, \ldots, s_d}$ exhibits a low-rank structure. 
Here, a low-rank structure refers to a property of the function whereby accuracy can be maintained even when the bond dimensions are bounded 
by some constant $\chi$.
The tilde in $\tilde{F}_{s_1 s_2\dotsc s_d}$ indicates TT compression of the target tensors; this notation will be used hereafter.

In the same manner, the $2d$-way tensor $F_{s_1, \dots, s_d}^{t_1, \dots, t_d} \in \mathbb{R}^{N_1 \times M_1 \times \cdots \times N_d \times M_d}$, where each local index $s_l$ (resp. $t_l$) has local dimension $N_l$ (resp. $M_l$), 
can be expressed with the product of the each tensor core, i.e., fourth-way tensor
$[F^{(i)}]^{t_i}_{a_{i-1}\,s_i\,a_i}
\in \mathbb{R}^{\chi_{i-1} \times N_i \times M_i \times \chi_i}$ as follows:
\begin{align}
    F_{s_1, \ldots, s_d}^{t_1, \ldots, t_d} \approx
     \tilde{F}_{s_1, \ldots, s_d}^{{t_1, \ldots, t_d}}
     &=
    \sum_{a_1=1}^{\chi_1} 
    \cdots 
    \sum_{a_{d-1}=1}^{\chi_d} 
    [F^{(1)}]^{t_1}_{a_0 s_1 a_1}
    \cdots 
    [F^{(d)}]_{a_{d-1} s_d a_d}^{t_d}
    \equiv 
    \prod^{d}_{i} [F^{(i)}]_{s_i}^{t_i}.
\end{align}

This representation is referred to as the tensor train operator (TTO), or matrix product operator. 
In this article, we simply call it TT, together with the tenor train.
Figure~\ref{fig:tt_zu} (b) shows the corresponding diagram.

\subsection{Tensor cross interpolation}
Tensor cross interpolation (TCI) is an adaptive method for constructing a TT representation $\tilde{F}$ of a high-dimensional tensor $F \in \mathbb{R}^{N_1 \times \cdots \times N_d}$.
One can view each entry $F_{s_1,\dots,s_d} = F(s_1,\dots,s_d)$ as a function value on a structured grid.
Rather than accessing all the $\prod_{i=1}^{d} N_{i}$ entries, TCI adaptively queries only a small, carefully chosen subset of these entries according to a specific rule.

In TCI, we successively take entries (``pivot") in $F$ specified by multi-indices $(s_1, \dots, s_d)$.
Starting from an initial pivot, TCI adaptively searches for the entry $(s_1, \dots, s_d)$ where the absolute error $|F_{s_1, \dots, s_d} - \tilde{F}_{s_1, \dots, s_d}|$ is largest, calculates 
the corresponding function value, and updates the TT approximation. 
This procedure is repeated until the normalized maximum error defined by
\begin{align}
\epsilon_{\mathrm{TCI}}
= \frac{\|F - \tilde{F}\|_\infty}{\|F\|_\infty}
= \frac{\max_{s_1,\dots,s_d} \bigl|F_{s_1,\dots,s_d} - \tilde{F}_{s_1,\dots,s_d}\bigr|}
       {\max_{s_1,\dots,s_d} \bigl|F_{s_1,\dots,s_d}\bigr|}
\label{eq:epsilon_TCI}
\end{align}
falls below a user-specified tolerance $\epsilon_{\mathrm{TCI}}$.
Here, normalising by $\lVert F\rVert_\infty$ safeguards against numerical instabilities that arise when the largest entry of $F$ is small.
In practice, computing $\epsilon_{\mathrm{TCI}}$ over the entire index set is prohibitive. Instead, we approximate \eqref{eq:epsilon_TCI} using only the entries sampled so far and terminate the iteration when this estimate falls below the prescribed tolerance. 
If the target tensor has a large bond dimension, the index space to be explored grows proportionally. Consequently, the algorithm often becomes trapped in flat regions, local minima where the error can no longer be reduced. 
And thus its struggles to reach unexplored areas that may contain much larger errors.


The algorithm samples only $O(d\,\chi^2 N_{\max})$ entries, where $\chi$ is the maximum bond dimension determined by the tolerance, and $N_{\max} = \max_i N_i$.
The total complexity is $O(d\,\chi^3\,N_{\max})$, increasing bond dimension from 1 to $\chi$~\cite{N_ez_Fern_ndez_2025}.
In contrast, applying SVD to the uncompressed tensor necessitates full access to all $\prod_{i}^d N_i$ entries.
Through its adaptive calling of target tensors, TCI thus uncovers low-rank structure from very few samples, making it especially suitable for tensors with exponentially many entries, namely with larger $d$.

\subsection{Singular value decomposition}
In this study, we use SVD to compress further the given TTs (e.g., those obtained from TCI where the TCI threshold $\epsilon_{\mathrm{TCI}}$ is set to a sufficiently low value).
This is achieved by first canonicalizing the TT using QR decompositions from left to right, and then performing the compression via SVD from right to left, discarding any remaining singular values whose sum of square is smaller than the specified tolerance $\epsilon_{\mathrm{SVD}}$.
This tolerance is defined by
\begin{align}
\epsilon_{\mathrm{SVD}}
=
\frac{|\tilde{F}_{\mathrm{TT}}-\Tilde{F}^{'}_{\mathrm{TT}}|^2_{\mathrm{F}}}{|\tilde{F}_{\mathrm{TT}}|^2_{\mathrm{F}}},
\label{eq:epsilon_SVD}
\end{align}
where $|\cdots|_{\mathrm{F}}$ indicates the Frobenius norm, and $\tilde{F}_{\mathrm{TT}}$ and $\Tilde{F}^{'}_{\mathrm{TT}}$ are the TTs before and after SVD, respectively.
The total complexity is $O(d\,\chi^3\,N_{\max})$~\cite{N_ez_Fern_ndez_2025}.
For more technical details, readers are referred to Ref.~\cite{Schollwock2011-eq, Oseledets2011}.
In the following, since we apply SVD to various target TTs, we define $\epsilon_{\mathrm{SVD}}(\cdot)$, where $\cdot$ denotes the target TT (e.g., $\tilde{F}_{\mathrm{TT}}$).

\section{Fourier-transform-based option pricing and Greeks computation}\label{sec:FT-basedpricing}
In this section, we present Fourier‐transform-based pricing of multi-asset options and calculation of associated Greeks under the BS model.
\subsection{Black-Scholes model}
In this article, we consider the underlying asset prices $\vec{S}(t)=(S_1(t),\cdots,S_d(t))  \in \mathbb{R}^d$ in the Black-Scholes (BS) model described by the following stochastic differential equation
\begin{align}
    dS_{m}(t) = rS_{m}(t)dt + \sigma_m S_{m}(t) dW_{m}(t),
\label{black and scholes multi assets}
\end{align}
where $r\in\mathbb{R}$ and $\sigma_1,\cdots,\sigma_d>0$ are constant parameters called the risk-free interest rate and the volatilities, respectively.
Here, $W_1(t),\cdots,W_d(t)$ are Brownian motions whose pairwise correlations are determined by a time-independent symmetric matrix \(\rho=(\rho_{mn})\), namely
\begin{align}
    dW_{m}(t) dW_{n}(t) = \rho_{mn}dt.
\label{gauss correlation}
\end{align}
The present time is set to $t=0$ and the present asset prices are denoted by $\vec{S}^0=(S^0_{1},\cdots,S^0_{d})$.

We consider European options whose payoff $v(\vec{S}(T))$ depends on the asset prices $\vec{S}(T)$ at the maturity $T$.
According to the theory of option pricing, the price $V$ of such an option is given by the expectation of the discounted payoff~\cite{hull2003options}:
\begin{align}
    V(\vec{\sigma}, \vec{S}^{0})&=\mathbb{E}\left[e^{-rT}v(\vec{S}(T))\middle|\vec{S}^0\right] \nonumber \\
    &=e^{-rT} \int_{-\infty}^{\infty} v(\exp(\vec{x}))q(\vec{x}|\vec{x}^0)dx,
    \label{expec val}
\end{align}
where we define $\exp(\vec{x}):=(e^{x_1},\cdots,e^{x_d})$.
$q(\vec{x}|\vec{x}^0)$ is the probability density function of $\vec{x}:=(\log S_1(T),\cdots,\log S_d(T))$, the log asset prices at $T$, conditioned on the present value $\vec{x}^0=(\log S^0_{1},\cdots,\log S^0_{d})$.
In the BS model defined by \eqref{black and scholes multi assets}, $q(\vec{x}|\vec{x}^0)$ is given by the $d$-variate normal distribution:
\begin{equation}
    q(\vec{x}|\vec{x}^0) = \frac{1}{\sqrt{(2 \pi)^d \det \Sigma}} \exp\left(-\frac{1}{2}\left(\vec{x}-\vec{\mu}\right)^T \Sigma^{-1} \left(\vec{x}-\vec{\mu}\right)\right),
\end{equation}
where $\Sigma_{mn}:=\sigma_m\sigma_n \rho_{mn}T$ is the covariance matrix of $\vec{x}$ and \\
$\vec{\mu}:=\vec{x}^0+\left(rT-\frac{1}{2}\sigma_1^2T,\cdots,rT-\frac{1}{2}\sigma_d^2T\right)$.

Note that in Eq.~\eqref{expec val}, the option price is denoted by $V(\vec{\sigma}, \vec{S}^{0})$, highlighting its dependence on the input parameter $\vec{\sigma}, \vec{S}^{0}$. 

\subsection{Fourier transform-based pricing}
In FT-based option pricing, we reformulate the expectation formula \eqref{expec val} as an integral in the Fourier space:
\begin{equation}
V(\vec{\sigma}, \vec{S}^0) = \frac{e^{-rT}}{(2\pi)^d} \int_{\mathbb{R}^d}
\varphi(-\vec{z}-i\vec{\alpha}, \vec{\sigma}, \vec{S}^0)
\tilde{v}(\vec{z}+i\vec{\alpha})d\vec{z},
\label{eq:VFTBased}
\end{equation}
where $\vec{z}\in\mathbb{R}^d$ denotes the Fourier variable corresponding to the $\vec{x}$ in the Fourier transform.
The vector $\vec{\alpha}\in\mathbb{R}^d$ is introduced to shift the integration contour into the complex domain.
\begin{equation}
\varphi(\vec{z}, \vec{\sigma}, \vec{S}^0):=\mathbb{E}[e^{i\vec{z}\cdot\vec{x}}|\vec{x}^0]=\int_{\mathbb{R}^d} e^{i\vec{z}\cdot\vec{x}}q(\vec{x}|\vec{x}^0)d\vec{x}    
\end{equation}
is the characteristic function, and in the BS model, it is given by~\cite{Schmelzle2010OptionPF, Eberlein16062010}
\begin{align}
    \varphi(\vec{z}, \vec{\sigma}, \vec{S}^0) = \exp \left(i \sum_{m=1}^d z_m \mu_m-
    \frac{T}{2} 
    \sum_{m=1}^d 
    \sum_{k=1}^d 
    \sigma_m \sigma_k z_m z_k \rho_{m k}\right). 
\label{eq:phi_tt}
\end{align}
$\tilde{v}(\vec{z}):=\int_{\mathbb{R}^d} e^{i\vec{z}\cdot\vec{x}}v(\exp(\vec{x}))d\vec{x}$ is the Fourier transformed payoff function, and its explicit formula is known for some types of options.
For example, for a European min-call option with strike $K$, which we will consider in our numerical demonstration, the payoff function is
\begin{align}
    v^{\text{min}}(\vec{S}_T) = \max\{\min\{S_{1}(T), \ldots, S_{d}(T)\} - K, 0\},
\label{min option}
\end{align}
and its Fourier transformation is \cite{Eberlein2010}
\begin{align}
    \tilde{v}^{\min }(\vec{z} + i\vec{\alpha})=-\frac{K^{1+i \sum_{m=1}^d (z_m + i\alpha_m)}}{(-1)^d\left(1+i \sum_{m=1}^d (z_m+i\alpha_m)\right) \prod_{m=1}^d i (z_m + i\alpha_m)}.
\label{eq:vmin_tt}
\end{align}
Because the Fourier transform of the
payoﬀ function has a branch cut in the complex plane, we introduce $\vec{\alpha}$ with $\alpha_m>0$ and $\sum_{m=1}^d\alpha_m>1$ for $\tilde{v}^{\min}$ to be well-defined. 

In the numerical calculation of Eq.~\eqref{eq:VFTBased}, we truncate each dimension of the infinite integration domain to a finite range
$[-\vec{z}{^{\min}}, \vec{z}{^{\max}}$, neglecting the tail contributions. 
This yields the modified integral:
\begin{align}
V(\vec{\sigma}, \vec{S}^0) 
&\simeq 
 \frac{e^{-rT}}{(2\pi)^d} \int_{-z^{\min}_1}^{z^{\max}_1}\cdots \int_{-z^{\min}_d}^{z^{\max}_d}
\varphi\bigl(-\vec{z}- i \vec{\alpha}, \vec{\sigma}, \vec{S}^0\bigr) 
\tilde{v}^{\min}\bigl(\vec{z} + i\vec{\alpha}\bigr) d\vec{z}.
\label{eq:V_truncated}
\end{align}

We then discretize the truncated integral by discretizing the variable $\vec{z}$ over a finite grid as follows:
\begin{align}
V(\vec{\sigma}, \vec{S}^0) 
&\simeq 
 \frac{e^{-rT}}{(2\pi)^d} \sum_{j_1, \dots, j_d=1}^{N_{z}} \varphi\bigl(-\vec{z}_{\mathrm{gr},\vec{j}} - i \vec{\alpha}, \vec{\sigma}_{}, \vec{S}^0_{}\bigr) 
\tilde{v}^{\min}\bigl(\vec{z}_{\mathrm{gr},\vec{j}} + i\vec{\alpha}\bigr) \prod_{{i}=1}^d w^{(i)}_{j_i}, 
\label{eq:V_discretized}
\end{align}
where $N_{z}$ denotes the number of grid points in each dimension and $w^{(i)}_{j_i}$ denotes the quadrature weight (i.e., integration weight) for the \(i\)th dimension, associated with the grid points \({z}_{\mathrm{gr},{j_i}}\).
Section~\ref{sec:discretization} provides further details on these grid points and quadrature weight.

\subsection{Greeks of Fourier-transform-based pricing}
By analytically differentiating the FT-based option price, we can derive closed-form expressions for the Greeks.
Here, we assume that the interchange of the order of integration and differentiation is possible.
In this study, we focus on three types of Greeks with respect to the $\kappa$-th parameter $p_{\kappa} \in \{ \sigma_\kappa, S^0_\kappa \}$. 
After discretization of the integration, the $l$-th derivative of the option value with respect to  $p_{\kappa}$ is approximated by
\begin{equation}
  \frac{\partial^\ell V(\vec{\sigma}, \vec{S}^0)}{\partial (p_\kappa)^\ell}
  \;\approx\;
  \frac{e^{-rT}}{(2\pi)^d} 
  \sum_{j_1, \dots, j_d=1}^{N_{z}} 
  \frac{\partial^\ell\varphi\bigl(-\vec{z}_{\mathrm{gr},\vec{j}} - i \vec{\alpha}, \vec{\sigma}_{}, \vec{S}^0_{}\bigr)}{\partial (p_\kappa)^\ell}
\tilde{v}^{\min}\bigl(\vec{z}_{\mathrm{gr},\vec{j}} + i\vec{\alpha}\bigr) \prod_{{i}=1}^d w^{(i)}_{j_i}.
  \label{eq:general_form}
\end{equation}

The first derivative of $V$ with respect to $\sigma_\kappa$ is commonly called the Vega:
\begin{align}
 \nu_{\kappa}(\vec{\sigma}, \vec{S}^0)
  &\approx
  \frac{e^{-rT}}{(2\pi)^d} 
  \sum_{j_1, \dots, j_d=1}^{N_{z}} 
  \Psi_{\nu_\kappa}(-\vec{z}_{\mathrm{gr},\vec{j}} - i \vec{\alpha}, \vec{\sigma})
  {\varphi\bigl(-\vec{z}_{\mathrm{gr},\vec{j}} - i \vec{\alpha}, \vec{\sigma}_{}, \vec{S}^0_{}\bigr)} 
  \notag\\
  &\quad
  \times
\tilde{v}^{\min}\bigl(\vec{z}_{\mathrm{gr},\vec{j}} + i\vec{\alpha}\bigr) \prod_{{i}=1}^d w^{(i)}_{j_i},
  \label{eq:Vega}
\end{align}
where we define the following quantity related to Vega, 
\begin{equation}
  \Psi_{\nu_\kappa}(-\vec{z} - i \vec{\alpha}, \vec{\sigma})
  \;=\;
  T\bigl(z_\kappa + i\alpha_\kappa\bigr)
  \;\Bigl(
      i\,\sigma_{\kappa} 
      \;+\; 
      \sum_{i=1}^{d} \rho_{i \kappa}\,\sigma_i \,\bigl(-z_i - i\alpha_{i}\bigr)
  \Bigr).
  \label{eq:Vega_factor}
\end{equation}

The first derivative of $V$ with respect to the initial asset price $S^0_{\kappa}$ is the Delta:
\begin{align}
\Delta_{\kappa}(\vec{\sigma}, \vec{S}^0)
  &\approx
  \frac{e^{-rT}}{(2\pi)^d}
  \sum_{j_1, \dots, j_d = 1}^{N_{z}}
  \Psi_{\Delta}\!\bigl(-z_{\mathrm{gr}, j_{\kappa}} - i\alpha_{\kappa},\, S^0_{\kappa}\bigr)\,
  \varphi\!\bigl(-\vec{z}_{\mathrm{gr}, \vec{j}} - i\vec{\alpha},\, \vec{\sigma},\, \vec{S}^0\bigr)
  \notag\\
  &\quad
  \times \tilde{v}^{\min}\!\bigl(\vec{z}_{\mathrm{gr}, \vec{j}} + i\vec{\alpha}\bigr)\,
  \prod_{{i}=1}^d w^{(i)}_{j_i},
\label{eq:delta}
\end{align}
where we define the following quantity related to Delta, 
\begin{equation}
  \Psi_{\Delta}(-z_\kappa - i \alpha_\kappa, S^0_{\kappa})
  \;=\;
  i\,\frac{-z_\kappa - i \alpha_{_{\kappa}}}{S^0_\kappa}.
  \label{eq:delta_factor}
\end{equation}

The second derivative of $V$ with respect to $S^0_{\kappa}$ is the Gamma:
\begin{align}
\gamma_{\kappa}(\vec{\sigma}, \vec{S}^0)
  &\approx
  \frac{e^{-rT}}{(2\pi)^d} 
  \sum_{j_1, \dots, j_d=1}^{N_{z}}
   \Psi_{\gamma}\!\bigl(-z_{\mathrm{gr},j_\kappa} - i\alpha_{\kappa},\, S^0_{\kappa}\bigr)\,
   \varphi\!\bigl(-\vec{z}_{\mathrm{gr},\vec{j}} - i\vec{\alpha},\,\vec{\sigma},\,\vec{S}^0\bigr)
   \notag\\
  &\quad
   \times \tilde{v}^{\min}\!\bigl(\vec{z}_{\mathrm{gr},\vec{j}} + i\vec{\alpha}\bigr)\,
   \prod_{{i}=1}^d w^{(i)}_{j_i},
\label{eq:gamma}
\end{align}
where we define the following quantitiy related to Gamma, 
\begin{equation}
 \Psi_{\gamma}(-z_\kappa - i {\alpha_{\kappa}}, S^0_{\kappa})
  \;=\;
  \,\frac{i(z_\kappa + i\alpha_{\kappa}) - \bigl(z_\kappa + i\alpha_{\kappa}\bigr)^2}{\bigl(S^0_\kappa\bigr)^2}.
  \label{eq:gamma_factor}
\end{equation}

\section{TT representations of Greeks}\label{GreeksTT}
\begin{figure}[H]
    \centering
    \includegraphics[width=1.0\linewidth]{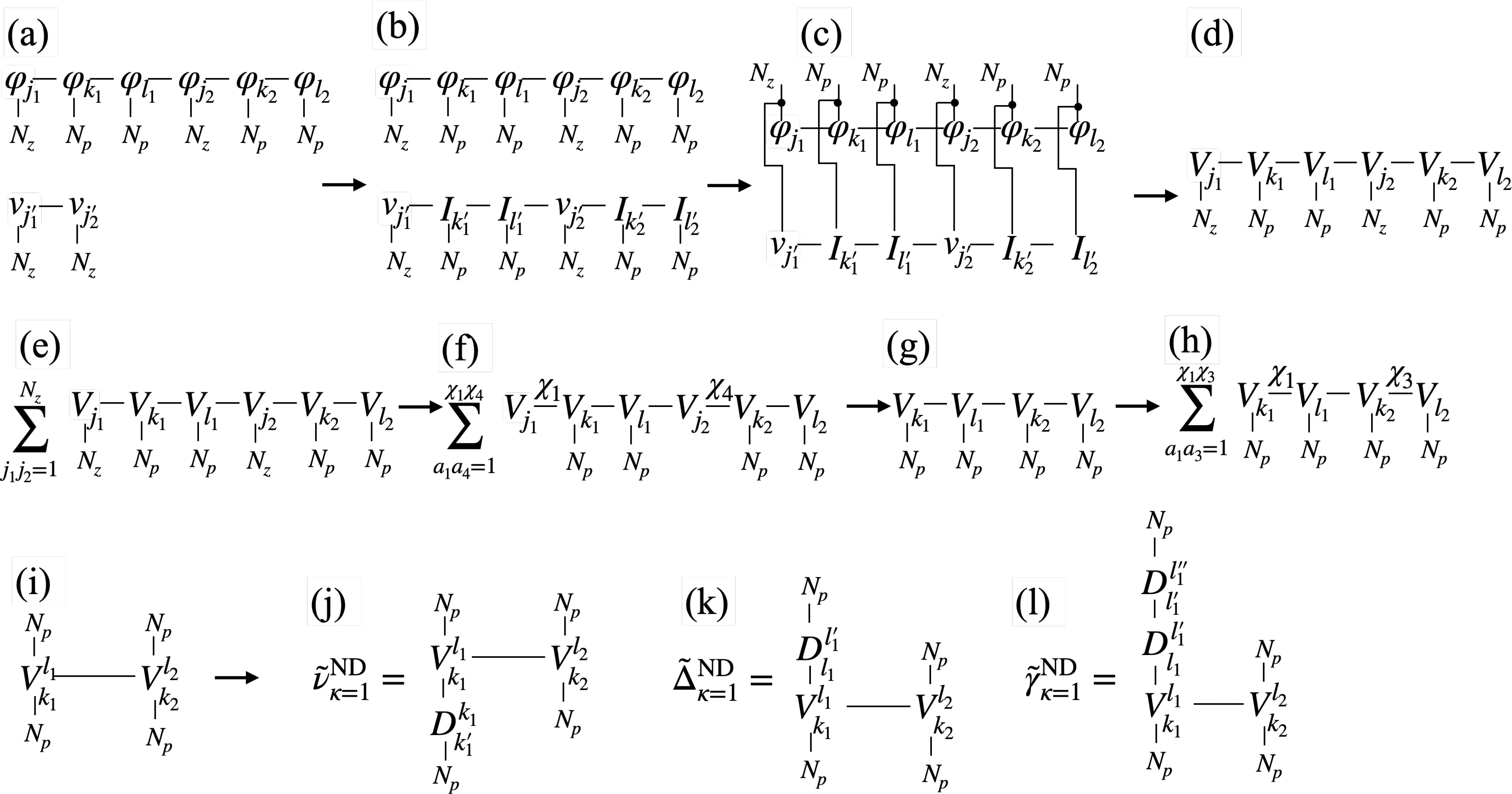}
    \caption{
    The procedure for constructing TT representations of the option price $V(\vec{p})$ and associated Greeks via the ND approach for $d=2$. 
    Panels (a)–(i) illustrate how to construct the TT representation of $V(\vec{p})$. 
    Panels (j)–(l) show the TT representations of Delta and Gamma with respect to $S_{\mathrm{gr},1}^0$, and Vega with respect to $\sigma_{\mathrm{gr},1}$.  
    }
    \label{fig:new_scheme_ND}
\end{figure}

We propose two TT-based methods for computing Greeks, which extend the previously proposed approach~\cite{math13111828}: (1) an numerical differentiation (ND) approach, 
and (2) an analytical differentiation (AN) approach.
Both TT-based methods share a common starting point, shown in steps (a)–(i) of Figure~\ref{fig:new_scheme_ND}, where a TT representation of the FT-based option price is built. 
From this point, the proposed method splits into two distinct approaches:
\begin{enumerate}
    \item[(1)] \textbf{Numerical differentiation approach}: 
    In steps (j)–(l) of Figure \ref{fig:new_scheme_ND}, we construct the TT representations of Greeks by applying numerical differentiatin operator(s) directly to the relevant core(s) of the TT representation of price.
    \item[(2)] \textbf{Analytical differentiation approach}: 
    In steps (m)–(o) of Figure \ref{fig:new_scheme_an}, we construct the TT representations of the Greeks derived from closed-form formulas for FT-based option pricing.
\end{enumerate}

\subsection{TT representation of Fourier transform-based option price}
Using TCI, we obtain the TTs to approximate $\varphi$ and $\tilde{v}$ that incorporate the dependency on the parameters $\vec{\sigma}$ and $\vec{S}^{0}$.
In other words, we view \(\varphi\) not just as a function of \(\vec{z}\) but also of \(\vec{\sigma}\) and \(\vec{S}^0\). 
We then set up a grid points $(\vec{z}_{\mathrm{gr},\vec{j}},\vec{\sigma}_{\mathrm{gr},\vec{k}}, \vec{S}^{0}_{\mathrm{gr},\vec{l}})$ in the space of ($\vec{z}$, $\vec{\sigma}, \vec{S}^{0}$) which are labeled by the index vectors $\vec{j}$, $\vec{k}$ and $\vec{l}$.
The number of grid points is denoted by $N_p$ for each of the parameter vectors $\vec{\sigma}$ and $\vec{S}^{0}$, while it is $N_z$ for $\vec{z}$, as defined previously. 
Thus, the full tensor grid consists of $N_z^d$ points in the $\vec{z}$-direction and $N_p^d$ points in each of the $\vec{\sigma}$- and $\vec{S}^0$-directions, resulting in a total of $N_z^d N_p^{2d}$ grid points. 
Each of these grid points is adaptively chosen by TCI from the full set of grid points $\{\vec{z}_{\mathrm{gr},\vec{j}}, \vec{\sigma}_{\mathrm{gr},\vec{k}}, \vec{S}^{0}_{\mathrm{gr},\vec{l}}\}$. 
The specific grids and integration weight $w^{(i)}_{j_i}$ used in this study are shown in Section~\ref{sec:discretization}. 

To manage the local indices of TT, we place the three core tensors corresponding to $\bigl(z_i, \sigma_i, S_i^0\bigr)$ for each asset $i$ contiguously in the TT ordering, denoted by $(z_i,\sigma_i, S^0_i)_{i=1}^d$.
That is, the sequence of TT cores is ($z_1, \sigma_1,  S^{0}_1 ,z_2, \sigma_2,  S^{0}_2, \cdots z_d, \sigma_d,  S^{0}_d$). 
Empirically, this interleaved arrangement (rather than separating the parameters, i.e., ($z_1 z_2 \cdots z_d \sigma_1 \sigma_2 \cdots \sigma_d  S^0_1 S^{0}_2 \cdots  S^{0}_d$) ) yields TTs with smaller bond dimensions.

As illustrated in Figure~\ref{fig:new_scheme_ND} (a), we apply TCI to construct the TT approximations $\tilde{\varphi}_{({j}_i,{k}_i,{l}_i)_{i=1}^{d}}$ (i.e., $\tilde{\varphi}_{j_1,k_1,l_1,\ldots,j_d,k_d,l_d}$) and $\tilde{v}^{\min}_{\vec{j}}$ (i.e., $\tilde{v}^{\min}_{j_1,\ldots,j_d}$), which respectively approximate the integrand $\tilde{\varphi}_{({j}_i,{k}_i,{l}_i)_{i=1}^{d}}$ weighted by a product of quadrature weights $\prod^{d}_{i=1} w^{(i)}_{j_i} $, and $\tilde{v}^{\min}_{j_1,\ldots,j_d}$.
Their entries correspond to the values of the respective function on the adaptive grid:
\begin{align}
 \prod^{d}_{i=1} w^{(i)}_{j_i} \times \varphi(\vec{z}_{\mathrm{gr},\vec{j}},\vec{\sigma}_{\mathrm{gr},\vec{k}}, \vec{S}^{0}_{\mathrm{gr},\vec{l}})
\simeq
\tilde{\varphi}_{({j}_i,{k}_i,{l}_i)_{i=1}^{d}}
 = \prod^{d}_{i=1} \varphi^{(3i-2)}_{j_i} \varphi^{(3i-1)}_{k_i} \varphi^{(3i)}_{l_i}, 
\label{eq:phi_tt_weighted}
\end{align}
and
\begin{align}
 \tilde{v}^{\mathrm{min}}(\vec{z}_{\mathrm{gr},\vec{j}})
  \simeq  \tilde{v}^{}_{\vec{j}}
 = \prod^{d}_{i} v^{(i)}_{j_i}.
\label{eq:v_tt_weighted}
\end{align}
evaluated at the grid points $(\vec{z}_{\mathrm{gr},\vec{j}},\vec{\sigma}_{\mathrm{gr},\vec{k}}, \vec{S}^{0}_{\mathrm{gr},\vec{l}})$ and $(\vec{z}_{\mathrm{gr},\vec{j}})$, respectively.
The TT representation of the product of integral weights has an exact bond dimension of one~\footnote{Alternatively, the integral weights can be incorporated into $\tilde{v}^{\min}$, or they can be directly applied during the summation (integration) step.}.
Finally, after running TCI with a sufficiently low tolerance, we reduce the bond dimensions of the resulting TTs by applying SVD, yielding a TT representation with lower bond dimensions.

To compute the elementwise multiplication of tensors $\tilde{\varphi}$ and $\tilde{v}$, the two tensors must first share a common dimensional structure.
For this purpose, we introduce dummy tensors $I^{(i)}_{a_i, s_i, b_i}$, whose role is to extend the dimensions of the payoff tensor $\tilde{v}$ to match the tensor structure of $\tilde{\varphi}$.
Each dummy tensor $I^{(i)}$ acts as an identity tensor along the bond indices, expanding $\tilde{v}$ into the same multi-index format as $\tilde{\varphi}$.
This allows the two tensors to have identical indexing structures, enabling straightforward elementwise multiplication while keeping the original payoff values.
Formally, for each $i \in \{1, \cdots d \}$, we define the dummy tensor as
\begin{align}
I^{(i)}_{a_i, s_{i}, b_{i}} 
= 
\begin{cases}
1, & \text{if } a_i = b_{i}, \\
0, & \text{otherwise},
\end{cases}
\label{eq:dummy_tensor}
\end{align}
where $a_i, b_i \in \{1, \dots, \chi_i\}$, 
$\chi_i$ is the $i$-th bond dimension of $\tilde{v}^{\mathrm{min}}$, 
and $s_{i} \in \{1, \dots, N_p\}$.
As illustrated in Figure.~\ref{fig:new_scheme_ND} (b), we insert these dummy tensors $I^{(i)}_{s_i}$ into TT at the positions corresponding to the $\vec{k}$ and $\vec{l}$-depdendent tensor cores in $\tilde{\varphi}_{({j}_i,{k}_i,{l}_i)_{i=1}^{d}}$.
This is given by
\begin{align}
\tilde{v}_{({j}_i,{k}_i,{l}_i)_{i=1}^{d}}
= 
\prod_{i=1}^{d}
{v}^{(i)}_{j_i} 
I^{(i)}_{k_i} 
I^{(i)}_{l_i} 
\label{eq:dummy_tensor_inseted}
\end{align}
where the $\tilde{v}$ now shares the same grid-indexing structure as $\tilde{\varphi}$.

In  Figure~\ref{fig:new_scheme_ND} (c), the elementwise multiplication of $\tilde{\varphi}$ and $\tilde{v}$ denoted by $\tilde{V}$, is implemented as
\begin{align}
& \tilde{V}_{(j_i,k_i,l_i)_{i=1}^{d}} \nonumber \\
&=
\sum_{\vec{j'}=1}^{N_z}
\sum_{\vec{k'}=1}^{N_p}
\sum_{\vec{l'}=1}^{N_p}
    (\prod^{d}_{i=1} \varphi^{(3i-2)}_{j_i} \varphi^{(3i-1)}_{k_i} \varphi^{(3i)}_{l_i})\;
   (\prod_{i=1}^{d}
   \delta_{j_i\,j'_i}\,
   \delta_{k_i\,k'_i}\,
   \delta_{l_i\,l'_i} )
   (\prod_{i=1}^{d}
{v}^{(i)}_{j'_i} 
I^{(i)}_{k'_i} 
I^{(i)}_{l'_i})
\nonumber \\
&= \sum_{\vec{j'}=1}^{N_z}
\sum_{\vec{k'}=1}^{N_p}
\sum_{\vec{l'}=1}^{N_p}
    (\prod^{d}_{i=1} \varphi^{(3i-2)}_{j_i} \delta_{j_i\,j'_i}
    \varphi^{(3i-1)}_{k_i} \delta_{k_i\,k'_i}
    \varphi^{(3i)}_{l_i} \delta_{l_i\,l'_i}) \;
   (\prod_{i=1}^{d}
{v}^{(i)}_{j'_i} 
I^{(i)}_{k'_i} 
I^{(i)}_{l'_i})
\nonumber \\
   &= 
   \sum_{\vec{j'}=1}^{N_z}
    \sum_{\vec{k'}=1}^{N_p}
    \sum_{\vec{l'}=1}^{N_p}
     (\prod^{d}_{i=1} [\varphi^{(3i-2)}]_{j_i}^{j'_i} 
     [\varphi^{(3i-1)}]_{k_i}^{j'_i} 
     [\varphi^{(3i)}]_{l_i}^{j'_i})
   (\prod_{i=1}^{d}
{v}^{(i)}_{j'_i} 
I^{(i)}_{k'_i} 
I^{(i)}_{l'_i})  \nonumber \\
   &= \prod_{i=1}^{d}  V^{(3i-2)}_{j_i} V^{(3i-1)}_{k_i} V^{(3i)}_{l_i},
\label{eq:elemntwise_phi_v}
\end{align}
where $\delta$  is a Kronecker delta function and $  [\varphi^{(i)}]_{j_i}^{j'_i} := \varphi^{(i)}_{j_i}\,\delta_{j_i j'_i}$  as shown in Figure~\ref {fig:new_scheme_ND} (d).
This produces a new TT, whose each bond dimension at the $i$-th bond equals the product of the corresponding bond dimensions of $\tilde{\varphi}$ and $\tilde{v}$. 
Concretely, if $\tilde{\varphi}$ and $\tilde{v}$ have bond dimensions $\chi_i$ and $\chi_i'$ at the $a_i$-th bond (for $i = 1, \dots, 3d-1$), then $\tilde{V}$ has a bond dimension $\chi_i \times \chi_i'$ at $a_i$-th bond. 

In Figure~\ref {fig:new_scheme_ND} (e), we sum over the $\vec{z}$ indices, an operation in TT, corresponds to multiplying each $\vec{j}$-dependent tensor core by a vector of ones~\footnote{
The procedure for constructing $\tilde{V}$ in this article differs from that described in~Ref.~\cite{math13111828}. 
Although the previous procedure, as in steps (3)-(4) in Figure 1 of Ref.~\cite{math13111828}, is conceptually straightforward, it suffers from the high computational complexity of the SVD performed immediately after the tensor contraction in step (3), particularly when correlations are random.
In contrast, the current method does not perform SVD immediately after elementwise multiplication. 
Instead, we first sum over the index $j$, significantly reducing the bond dimensions. 
This modification substantially lowers memory requirements and computational time.
}. 
Formally, 
\begin{equation}
\tilde{V}_{(k_i,l_i)_{i=1}^{d}} 
=
\sum^{N_{z}}_{\vec{j}=1}\prod_{i=1}^{d}  V^{(3i-2)}_{j_i} V^{(3i-1)}_{k_i} V^{(3i)}_{l_i} = 
\prod_{i=1}^{d}  V^{(3i-2)} V^{(3i-1)}_{k_i} V^{(3i)}_{l_i}.
\label{eq:sum_j}
\end{equation}

In Figure~\ref{fig:new_scheme_ND} (f), we contract adjacent tensor cores pairwise, specifically $V^{(3i-2)}_{{a_{3i-3} a_{3i-2}}}$ and $V^{(3i-1)}_{{a_{3i-2} k_{i} a_{3i-1}}}$ ($i=1, \cdots, d)$ as follows:
\begin{align}
V^{\prime(3i-1)}_{{a_{3i-3},k,a_{3i-1}}} 
=
\sum_{ a_{3i-2}=1}^{\chi_{3i-2}}
V^{(3i-2)}_{{a_{3i-3}, a_{3i-2}}} 
V^{(3i-1)}_{{a_{3i-2}, k_{i}, a_{3i-1}}}.
\end{align}
For clarity and simplicity of notation, we now redefine $V^{\prime(3i-1)}$ and $V^{3i}$ as $V^{(2i-1)}$ and $V^{(2i)}$, respectively.
We then optimize the bond dimensions of  $\tilde{V}_{(k_i,l_i)_{i=1}^{d}}$ by applying SVD as shown in Figure.~\ref{fig:new_scheme_ND} (g). 

Once the parameter indices $\vec{k}, \vec{l}$ are fixed for pricing and Greeks computations, the tensors no longer depend on the local parameter indices. To make this tensor contractions faster, we first contract adjacent tensor cores corresponding to indices $k_i$ and $l_i$, resulting in a new TT representation whose length is halved from $2d$ to $d$.
Note that, as we store this TT representation, the memory requirement increases from $O(2d \chi^2 N_p)$ to $O(d \chi^2 N_p^2)$.

As illustrated in Figure~\ref{fig:new_scheme_ND}(h), this procedure begins with the pairwise contraction of adjacent tensor cores, merging each pair of indices $(k_i, l_i)$.
\begin{align}
\tilde{V}_{(k_i,l_i)_{i=1}^{d}} 
&=
\sum^{\chi_{1}}_{a_{1}}\cdots
\sum^{\chi_{2d-1}}_{2d-1}
\tilde{V}^{\mathrm{}}_{k_1,l_1,\ldots, k_d,l_d} \nonumber \\
&= \sum^{\chi_{1}}_{a_{1}} \cdots \sum^{\chi_{2d-1}}_{a_{2d-1}}
V^{(1)}_{a_0 k_1 a_1} V^{(1)}_{a_1 l_1 a_2}
\cdots
V^{(d)}_{a_{2d-2} k_d a_{2d-1}} V^{(d)}_{a_{2d-1} l_d a_{2d}}.
\end{align}
We then reshape each tensor core of the resultant TT, $\tilde{V}_{(k_i,l_i)_{i=1}^{d}} $ into a 4-way tensor, $\tilde{V}^{k_1  \cdots k_d}_{l_1 \cdots l_d} =\tilde{V}^{\vec{l}}_{\vec{k}}=\prod^{d}_{i} 
[V^{k_i}_{l_i}]^{(i)}$
which produces the TT as illustrated in Figure.~\ref{fig:new_scheme_ND} (i).
With this TT, we can obtain the option price for the specified value of $(\vec{\sigma}, \vec{S}^{0})$.
By fixing the index $(\vec{k}, \vec{l})$ of $\tilde{V}^{\; k_1 k_2 \cdots k_d}_{\;\;l_1 l_2 \cdots l_d} $ 
to the value corresponding to the specified $(\vec{\sigma}, \vec{S}^{0})$, we get the option price for it as
\begin{equation}
     V(\vec{\sigma}_{\mathrm{gr},\vec{{k}}},
     \vec{S}^{0}_{\mathrm{gr},\vec{{l}}})
     \simeq
     \tilde{V}(\vec{\sigma}_{\mathrm{gr},\vec{{k}}},
     \vec{S}^{0}_{\mathrm{gr},\vec{{l}}})
     =
     \frac{e^{-rT}}{(2\pi)^d} 
      \tilde{V}^{\; k_1 k_2 \cdots k_d}_{\;\; l_1 l_2 \cdots l_d}.
\label{eq:option_price_TTO}
\end{equation}


\subsection{TT representation of Greeks based on numerical differentiation}
In the first approach, which we call the ND approach, as illustrated in Figure.~\ref{fig:new_scheme_ND} (j)-(l), we use numerical differentiation to efficiently compute Greeks with the TT representation.
To compute a particular Greeks, we apply a numerical differential operator $ D $ to the relevant index (either $ k_{\kappa} $ for volatility $\sigma_\kappa$ or $ l_{\kappa} $ for the underlying asset price $ S^0_\kappa $) in the $\kappa$-th TT core.
The explicit form of operator $D$ depends on how to discretize the parameter space. 
The specific discretization is given in Section~\ref{sec:discretization}.

As illustrated in Figure.~\ref{fig:new_scheme_ND} (j), (k), we obtain TT representation of first-order derivatives by applying the numerical differential operator to the $\kappa$-th TT core:
\begin{align}
\frac{\partial V(\vec{\sigma}_{\mathrm{gr},\vec{{k}}}, \vec{S}^{0}_{\mathrm{gr},\vec{{k}}})}{\partial \sigma_{\kappa}}
&\approx 
\tilde{\nu}^{\mathrm{ND}}_{\kappa}(\vec{\sigma}_{\mathrm{gr},\vec{{k}}}, \vec{S}^{0}_{\mathrm{gr},\vec{{k}}})  \nonumber  \\
&= \frac{e^{-rT}}{(2\pi)^d}
\, [V^{(1)}]^{k_1}_{l_1}
\cdot (\cdots) \cdot
\sum_{k_\kappa'=1}^{N_p}
D_{k_\kappa, k_\kappa'}^{(\kappa)} [V^{(\kappa)}]^{k_\kappa'}_{l_\kappa} 
\cdot (\cdots) \cdot
[V^{(d)}]^{k_d}_{l_d},
\end{align}

\begin{align}
\frac{\partial V(\vec{\sigma}_{\mathrm{gr},\vec{{k}}}, \vec{S}^{0}_{\mathrm{gr},\vec{{k}}})}{\partial S^0_{\kappa}}
&\approx 
\tilde{\Delta}^{\mathrm{ND}}_{\kappa}(\vec{\sigma}_{\mathrm{gr},\vec{{k}}}, \vec{S}^{0}_{\mathrm{gr},\vec{{k}}})  \nonumber  \\
&= \frac{e^{-rT}}{(2\pi)^d} 
\,[V^{(1)}]^{k_1}_{l_1}
\cdot (\cdots) \cdot
\sum_{l_\kappa'=1}^{N_p}
D_{l_\kappa, l_\kappa'}^{(\kappa)} \,[V^{(\kappa)}]^{k_\kappa}_{l_\kappa'}
\cdot (\cdots) \cdot
[V^{(d)}]^{k_d}_{l_d}.
\end{align}

For second-order derivatives with respect to $ S^0_{l_{\kappa}} $, we apply the first-order differentiation operator twice to the $\kappa$-th TT core, as illustrated in Figure.~\ref{fig:new_scheme_ND} (l):
\begin{align}
\frac{\partial^2 V(\vec{\sigma}, \vec{S}^{0})}{\partial^2 S^0_{\kappa}}
&\approx \tilde{\gamma}^{\mathrm{ND}}_{\kappa}(\vec{\sigma}, \vec{S}^{0})  \nonumber \\
&= \frac{e^{-rT}}{(2\pi)^d} 
\, [V^{(1)}]^{k_1}_{l_1}
\cdot (\cdots) \cdot
\sum_{l_\kappa'=1}^{N_p}
\sum_{l_\kappa''=1}^{N_p}
D_{l_\kappa, l_\kappa''}^{(\kappa)} 
\Bigl[D_{l_\kappa'', l_\kappa'}^{(\kappa)} [V^{(\kappa)}]^{k_\kappa}_{l_\kappa'}\Bigr]
\cdot (\cdots) \cdot
[V^{(d)}]^{k_d}_{l_d}.
\end{align}

Note that the differential operator $D$ used here is local, meaning it acts only on a single tensor core of the TT $\tilde{V}$.
As a result, the bond dimensions of the TT representations, $\tilde{\nu}^{\mathrm{ND}}$, $\tilde{\Delta}^{\mathrm{ND}}$, and $\tilde{\gamma}^{\mathrm{ND}}$ remain the same as those of the original tensor $\tilde{V}$.
This is beneficial because calculating the Greeks using their TT representations does not increase computational cost beyond that required for the price evaluation itself.

\begin{figure}[H]
    \centering
    \includegraphics[width=1.0\linewidth]{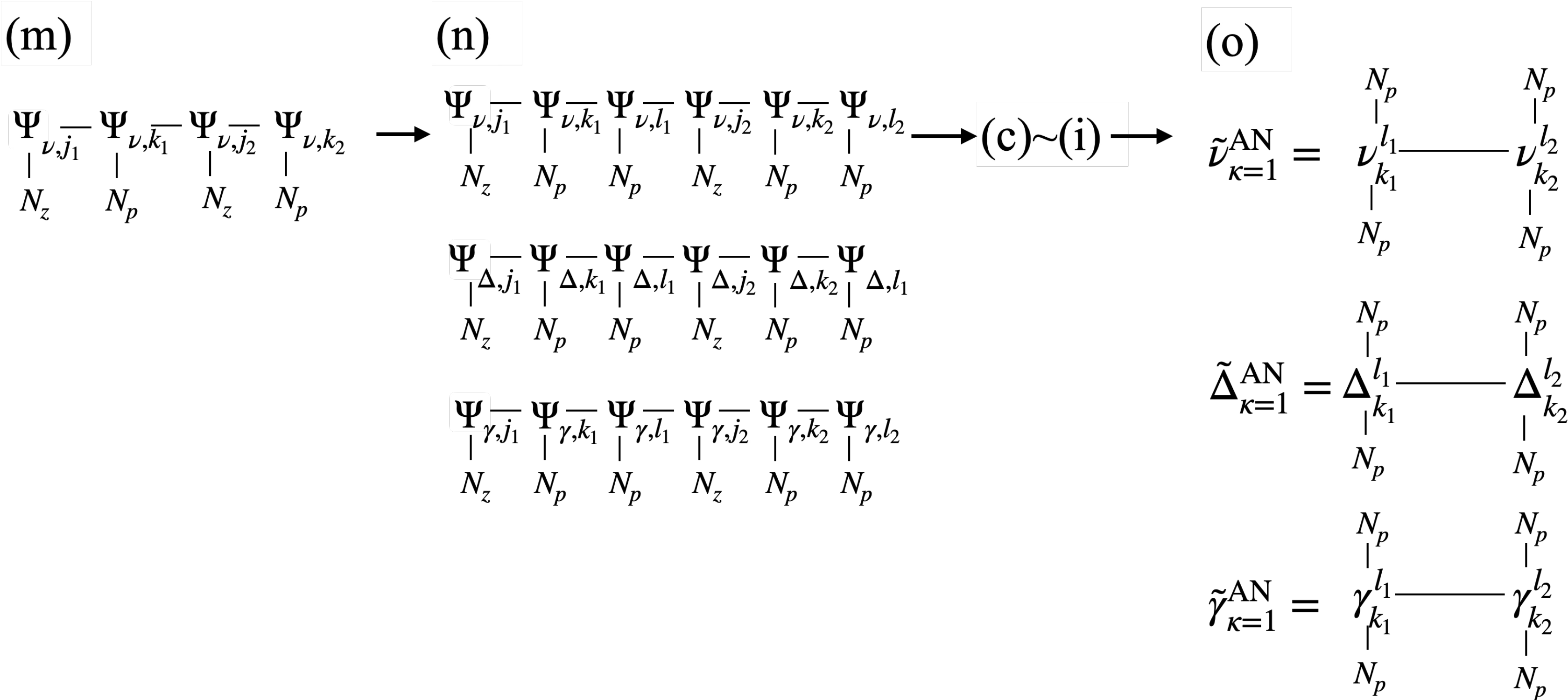}
    \caption{
   The procedure for constructing TT representations of Greeks obtained via analytical differentiation of $V(\vec{p})$. Panel (m) and (n) show the TT form of the Greeks coefficinet (or price function).  Panel (o) illustrates the resulting TT representations for Delta and Gamma with respect to $S_{\mathrm{gr},1}^0$, and Vega with respect to $\sigma_{\mathrm{gr},1}$.  
For further details, refer to the main text.
    }
    \label{fig:new_scheme_an}
\end{figure}

\subsection{TT representations of Greeks based on analytical differentiation}
In the second approach, the AN approach, as illustrated in Figure.~\ref{fig:new_scheme_an} (m)-(o), we use analytical differentiation forms to compute the Greeks with TT representations.

First, we apply TCI to $\Psi_{\nu_\kappa}(\vec{z}_{\mathrm{gr}, \vec{j}}, \vec{\sigma}_{\mathrm{gr}, \vec{k}})$ as shown in Eq.~\eqref{eq:Vega_factor} to obtain a TT representation as
\begin{align}
\Psi_{\nu_\kappa}(\vec{z}_{\mathrm{gr}, \vec{j}}, \vec{\sigma}_{\mathrm{gr}, \vec{k}})
\simeq
 \tilde{\Psi}_{\nu_\kappa}(\vec{z}_{\mathrm{gr}, \vec{j}}, \vec{\sigma}_{\mathrm{gr}, \vec{k}})
 = \prod^{d}_{i=1} \Psi^{(3i-2)}_{\nu_\kappa, j_i} \Psi^{(3i-1)}_{\nu_\kappa, k_i}.
\label{eq:tt_phi}
\end{align}
Then, we insert dummy three-way tensors into TT at the positions corresponding to the $l$-dependence tensor cores, as illustrated in Figure~\ref{fig:new_scheme_an}(n), so that we can take element-wise multiplication with $\tilde{V}$.
This is given by: 
\begin{align}
 \tilde{\Psi}_{\nu_\kappa, (j_i, k_i, l_i)_{i=1}^{d}}
 = \prod^{d}_{i} \Psi^{(3i-2)}_{\nu_\kappa, j_i} 
 \Psi^{(3i-1)}_{\nu_\kappa, k_i} 
  \Psi^{(3i)}_{\nu_\kappa, l_i}.
\label{eq:tt_phi_insert_dummy}
\end{align}

As for $\Psi_{\Delta}(-\vec{z}_{\mathrm{gr},\kappa}, S^{0}_{\kappa})$, we do not need to use TCI here because $\Psi_{\Delta}$ depends only on $z_\kappa$ and $S^0_\kappa$ through a direct product form, as shown in Eq.~\eqref{eq:delta_factor}.
Therefore, we construct a TT whose $\kappa$-th core corresponds explicitly to the values of $\Psi_{\Delta}(-\vec{z}_{\mathrm{gr},\kappa}, S^{0}_{\kappa})$. At all other core positions, dummy tensors with bond dimension equal to one are inserted, as illustrated in Figure~\ref{fig:new_scheme_an}(n).
\begin{align}
\tilde{\Psi}_{\Delta_{\kappa}, (j_i, k_i, l_i)_{i=1}^{d}}=
 \prod^{d}_{i} \Psi^{(3i-2)}_{\Delta_{\kappa}, j_i} 
 \Psi^{(3i-1)}_{\Delta_{\kappa}, k_i} 
  \Psi^{(3i)}_{\Delta_{\kappa}, l_i}.
\label{eq:tt_D}
\end{align}

Similarly, $\Psi_{\gamma}$ defined in Eq.\eqref{eq:gamma_factor} can be written as a TT with bond dimension exactly one  as illustrated in Fig.~\ref{fig:new_scheme_an}(n) by applying the same procedure used for Eq.\eqref{eq:tt_D}.
\begin{align}
\tilde{\Psi}_{\gamma_{\kappa}, (j_i, k_i, l_i)_{i=1}^{d}}=
 \prod^{d}_{i} \Psi^{(3i-2)}_{\gamma_{\kappa}, j_i} 
 \Psi^{(3i-1)}_{\gamma_{\kappa}, k_i} 
  \Psi^{(3i)}_{\gamma_{\kappa}, l_i}.
\label{eq:tt_gamma}
\end{align}

We perform SVD on $\tilde{V}_{(j_i,k_i,l_i)_{i=1}^{d}}$ in Eq.~\eqref{eq:elemntwise_phi_v} to reduce its bond dimensions.
This process is used only in the AN approach. 
The purpose of this compression is to reduce the computational complexity of the subsequent element-wise multiplication and the following SVD.

The remaining steps essentially follow those illustrated in Figures~\ref{fig:new_scheme_ND}(c)–(i).
We first obtain TT representations for Vega, Delta, and Gamma with respect to the $\kappa$-th parameter, from Eqs.~\eqref{eq:tt_phi_insert_dummy}, \eqref{eq:tt_D}, and \eqref{eq:tt_gamma}. 
We then perform element-wise multiplication between each of these TT representations and the TT representation $\tilde{V}_{(j''_i,k''_i,l''_i)_{i=1}^{d}}$ from Eq.~\eqref{eq:elemntwise_phi_v}.
We denote the resulting element-wise multiplications as $\tilde{\nu}_{\kappa,(j_i,k_i,l_i)_{i=1}^{d}}$, $\tilde{\Delta}_{\kappa,(j_i,k_i,l_i)_{i=1}^{d}}$, and $\tilde{\gamma}_{\kappa,(j_i,k_i,l_i)_{i=1}^{d}}$.

We sum over the index $\vec{j}$ of each tensor, resulting in $\tilde{\nu}_{\kappa,(k_i,l_i)_{i=1}^{d}}$, $\tilde{\Delta}_{\kappa,(k_i,l_i)_{i=1}^{d}}$, and $\tilde{\gamma}_{\kappa,(k_i,l_i)_{i=1}^{d}}$.
Next, we compress the resulting TTs by reducing bond dimensions via SVD, followed by contracting adjacent tensor cores associated with indices $k$ and $l$.

Finally, as illustrated in Figure~\ref{fig:new_scheme_an}(o), we arrive at the TT representations of Vega, Delta, and Gamma:
\begin{align}
 \nu^{\mathrm{AN}}_{\kappa}(\vec{\sigma}_{\mathrm{gr}, \vec{k}}, \vec{S}^{0}_{\mathrm{gr}, \vec{l}}) 
 \simeq 
 \tilde{\nu}^{\mathrm{AN}}_{\kappa}(\vec{\sigma}_{\mathrm{gr}, \vec{k}}, \vec{S}^{0}_{\mathrm{gr}, \vec{l}})
=\frac{e^{-rT}}{(2\pi)^d}  \prod^{d}_{i} [\nu^{(i)}]_{l_i}^{k_i},
\end{align}

\begin{align}
 \Delta^{\mathrm{AN}}_{\kappa}(\vec{\sigma}_{\mathrm{gr}, \vec{k}}, \vec{S}^{0}_{\mathrm{gr}, \vec{l}}) 
 \approx
 \tilde{\Delta}^{\mathrm{AN}}_{\kappa}(\vec{\sigma}_{\mathrm{gr}, \vec{k}}, \vec{S}^{0}_{\mathrm{gr}, \vec{l}})
 =
 \frac{e^{-rT}}{(2\pi)^d}  \prod^{d}_{i} [\Delta^{(i)}]_{l_i}^{k_i},
\end{align}

\begin{align}
 \gamma^{\mathrm{AN}}_{\kappa}(\vec{\sigma}_{\mathrm{gr}, \vec{k}}, \vec{S}^{0}_{\mathrm{gr}, \vec{l}})
 \approx
 \tilde{\gamma}^{\mathrm{AN}}_{\kappa}(\vec{\sigma}_{\mathrm{gr}, \vec{k}}, \vec{S}^{0}_{\mathrm{gr}, \vec{l}})
  =\frac{e^{-rT}}{(2\pi)^d}  \prod^{d}_{i} [\gamma^{(i)}]_{l_i}^{k_i}.
\end{align}

Similar to Eq.~\eqref {eq:option_price_TTO}, we obtain TT representations of the Greeks by fixing the parameter indices at their desired values.

Because terms in the summations for Greeks (Eqs. \eqref{eq:Vega}, \eqref{eq:delta}, and \eqref{eq:gamma}) are equal to those in the summation for the option price (Eq. \eqref{eq:V_discretized}) multiplied by the factors $\Psi_{\nu_\kappa}$, $\Psi_\Delta$, and $\Psi_\gamma$, 
the AN approach possibly yields TT representations of Greeks whose bond dimensions exceed those of the TT of the option price. 
Indeed, as shown in the right panel of Fig. \ref{fig:bonddim}, our simulations below clearly show the increases of the bond-dimension for all three Greeks.


\subsection{Computational complexity}
In our numerical experiments below, having the TT representations, we focus on the computational cost of TT contraction, i.e., evaluating the Greeks during the online phase, shown in Figure~\ref{fig:new_scheme_an} (j)–(l) and Figure~\ref{fig:new_scheme_ND} (o).
In other words, the computational complexity and time reported in this article exclude the computational complexity of constructing TT representations, the offline-phase procedures illustrated in Figure~\ref{fig:new_scheme_an}(a)–(i) and Figure~\ref{fig:new_scheme_ND}(a)–(i).

The computational complexity of the online phase of the TT-based Greeks computation, which is evaluating a specific tensor component $\tilde{V}_{k_1,\ldots,k_d}$ for fixed $k_1,\ldots,k_d$, is $O(d \chi_{\tilde{V}}^{2})$.
Here, we denote the maximum bond dimensions of the TT by $\chi_{\tilde{V}}$.
In practice, the bond dimension can vary across different bonds, and accounting for this variation is crucial to accurately assess the total number of operations. 
We will consider this aspect when we evaluate the computational complexity of the TT-based method in our numerical experiment, which is summarized in Table~\ref{table:table_res}.

We now compare the offline computational complexity between the ND and AN approaches. 
The AN approach requires constructing the TT representations of the additional quantities, $\Psi_{\nu}$ for Vega, $\Psi_{\Delta}$ for Delta, and $\Psi_{\gamma}$ for Gamma.
In contrast, the ND approach only needs to apply the numerical differentiation operator directly to one of the tensor cores of the TT representation of price, resulting in lower computational complexity.
Therefore, the ND approach has an advantage in the offline phase.

\subsection{Monte Carlo-based Greeks computations}
Here, we also make a brief description of the MC-based option pricing and Greeks computation.
It 
serves as a benchmark in our numerical demonstration of the TT-based method.

In the MC-based approach, we estimate the option price, which is given as the expectation in Eq. \eqref{expec val}, by the average of the payoffs in the sample paths:
\begin{equation}
    V^{\mathrm{MC}}(\vec{\sigma}, \vec{S}^{0})\approx e^{-rT} \times \frac{1}{N_{\rm path}} \sum_{i=1}^{N_{\rm path}} v\left(\exp(\vec{x}_i)\right),
\end{equation}
where $\vec{x}_1,\ldots,\vec{x}_{N_{\rm path}}$ are i.i.d. samples from $q(\vec{x}|\vec{x}_0)$.
On how to sample multivariate normal variables, we leave the detail to textbooks (e.g., Ref. \cite{glasserman2004monte}) and just mention that it requires more complicated operations than simple multiplications and additions, e.g., pseudorandom number generation and evaluation of some elementary functions.
Besides, calculating the payoff $v$ with the normal variable $\vec{x}_i$ involves exponentiation.
In the MC simulation for $d$ assets with $N_{\rm path}$, the number of such operations is $O(d N_{\rm path})$, and we hereafter estimate the computational complexity of MC-based option pricing by this.

In this study, two- and three-point central finite differences are employed to approximate first- and second-order Greeks with second-order accuracy in the perturbation step size $h$~\cite{glasserman2004monte}. Specifically, the first derivative is approximated by

\begin{equation}
\frac{V^{\mathrm{MC}}(p_0 + h) - V^{\mathrm{MC}}(p_0 - h)}{2h},
\end{equation}
and the second derivative by
\begin{equation}
\frac{V^{\mathrm{MC}}(p_0 - h) - 2V^{\mathrm{MC}}(p_0) + V^{\mathrm{MC}}(p_0 + h)}{h^2}.
\end{equation}
Here, we fix the random seed across the MC simulations. 
Then, the biases of these estimators are $O(h^2)$, while the variances scale as $O(1)$ for the first derivative and $O(h^{-1})$ for the second derivative. 
Reducing $h$ excessively to suppress bias inevitably amplifies variance for the second-order derivative, requiring a significant increase in the number of simulation paths, $N_{\mathrm{path}}$, to maintain sufficient accuracy. 
Consequently, choosing an overly small $h$ leads to greater computational complexity.

\subsection{Malliavin calculus-based Greeks computations}
An alternative is to use Malliavin calculus, by which each of Greeks is expressed as a single expectation.
This method circumvents the bias due to discretization, and the variance, the only error source, converges by increasing the MC sample size.
Here, we merely summarise the resulting formulae taken from \cite{XU2014493}.

Let \({B} = \mathrm{diag}(\sigma_{1},\dots,\sigma_{d})\) be the diagonal volatility matrix and the Cholesky decomposition of the correlation matrix be \(\rho=R {R}^\mathsf{T}\). 
We define
\begin{equation}
    G \;=\; B\,R.
\end{equation}
For each path labeled by \(j=1,\dots,N_{\mathrm{Path}}\), we define the following quantities:
  \begin{equation}
    y_i^{(j)} 
    \;=\; 
    \ln\!\Bigl(\tfrac{S_i^{(j)}(T)}{S_{0,i}}\Bigr)
    \;-\;
    \Bigl(r - \tfrac12\sum_{k=1}^d G_{i,k}^2\Bigr)\,T,
  \end{equation}
and
  \begin{equation}
    U_T^{(j)}
    \;=\;
    {G}^{-1}\,\mathbf{y}^{(j)}
    \;=\;
    \bigl(U_T^{(j),1},\,\dots,\,U_T^{(j),d}\bigr)^\mathsf{T}.
  \end{equation}
Then, the MV-based formulas for the Vega $\nu^{\mathrm{MV}}_{\kappa}$, Delta $\Delta^{\mathrm{MV}}_{\kappa}$, and Gamma $\gamma^{\mathrm{MV}}_{\kappa}$ with respect to $\kappa$-th parameter are as follows: 
\begin{equation}
  \nu^{\mathrm{MV}}_{\kappa}
  \;=\;
  e^{-rT}\,\frac{G_{\kappa,\kappa}}{\sigma_\kappa}
  \;\frac{1}{N_{\mathrm{Path}}}
  \sum_{j=1}^{N_{\mathrm{Path}}}
  \Bigl[ v\left(\exp(\vec{x}_j)\right) \;\times\; \zeta_{(j)}\Bigr],
\end{equation}
where 
\begin{equation}
  \mathrm{\zeta}_{(j)}
  \;=\;
  \Bigl(\sum_{i=1}^d (G^{-1})_{i,\kappa}\,U_T^{(j),i}\Bigr)
  \Bigl(\tfrac{U_T^{(j),\kappa}}{T} \;-\; G_{\kappa,\kappa}\Bigr)
  \;-\;
  ({G}^{-1})_{\kappa,\kappa}.
\end{equation}

\begin{equation}
  \Delta^{\mathrm{MV}}_{\kappa}
  \;=\;
  \frac{e^{-rT}}{T\,S_{0,\kappa}}
  \;\frac{1}{N_{\mathrm{Path}}}
  \sum_{j=1}^{N_{\mathrm{Path}}}
  \Bigl[v\left(\exp(\vec{x}_j)\right) \;\times\; \theta_{(j)}\Bigr],
\end{equation}
where 
\begin{equation}
  \theta_{(j)}
  \;=\;
  \sum_{i=1}^d (G^{-1})_{i,\kappa}\,U_T^{(j),i}.
\end{equation}

\begin{equation}
  \gamma^{\mathrm{MV}}_{\kappa}
  \;=\;
  \frac{e^{-rT}}{T\,S_{0,\kappa}}
  \;\frac{1}{N_{\mathrm{Path}}}
  \sum_{j=1}^{N_{\mathrm{Path}}}
  \Bigl[v\left(\exp(\vec{x}_j)\right)\;\times\; \eta_{(j)}\Bigr],
\end{equation}
where 
\begin{equation}
  \eta_{(j)}
  = 
  -\frac{1}{S_{0,\kappa}}
  \sum_{i=1}^d ({G}^{-1})_{i,\kappa}\,U_T^{(j),i}
  \;+\;
  \frac{1}{S_{0,\kappa}\,T}
  \Bigl[
    \Bigl(\sum_{k=1}^d ({G}^{-1})_{k,\kappa}\,U_T^{(j),k}\Bigr)^2
    \;-\;
    T \sum_{k=1}^d ({G}^{-1})_{k,\kappa}^2
  \Bigr].
\end{equation}


\section{Numerical details}\label{numericaldetails}
We summarize the numerical settings employed in the numerical demonstration of the proposed methods.
\subsection{Discretization}\label{sec:discretization}
We employ the Gauss–Kronrod quadrature rule in discretizing integration variables $\vec{z}=(z_1, z_2, \cdots z_{d})$, which lie in a $d$-dimensional hyperrectangular domain 
\begin{equation}
\mathcal{X} 
\;=\;
[z_1^{\min}, z_1^{\max}] 
\,\times\,
\cdots
\,\times\,
[z_d^{\min}, z_d^{\max}]
\;\subset\;\mathbb{R}^{d}.
\end{equation}
Let $N_z$ is the number of grid points in each dimension. 
We set the gird points $\vec{z}_{\mathrm{gr},\vec{j}} $ to the Gauss–Kronrod nodes~\cite{etna_vol45_pp371-404}, for which the volume element $\prod_{{i}=1}^d w^{(i)}_{j_i}$ is given by the product of the Gauss–Kronrod weights\footnote{Although Gauss–Hermite quadrature can handle infinite domains, our numerical experiments showed that Gauss–Kronrod offered better accuracy. }.

We consider the parameter vector $\vec{p}=(\sigma_1, S^{0}_1, \cdots \sigma_d, S^0_{d})$ in a $2d$-dimensional hyperrectangular domain:
\begin{equation}
\mathcal{X} 
\;=\;
[\sigma_1^{\min}, \sigma_1^{\max}] 
\,\times\,
[S_1^{0,\min}, S_1^{0,\max}] 
\,\times\,
\cdots
\,\times\,
[\sigma_d^{\min}, \sigma_d^{\max}]
\,\times\,
[S_d^{0,\min}, S_d^{0,\max}]
\;\subset\;\mathbb{R}^{2d}.
\end{equation}
We discretize this domain with the tensor-product grid of the Chebyshev--Lobatto nodes~\cite{boyd01}. 
The one-dimensional Chebyshev–Lobatto nodes are defined on the interval $[-1,1]$ as
\begin{align}
L_{k}=\cos\!\Bigl(\tfrac{\pi\,(k-1)}{N_p-1}\Bigr),
\qquad k=1,\dots,N_p
\end{align}
with an integer $N_p \ge 2$.
For each coordinate $i = 1,\dots,2d$, we take $N_p$ points, $p_{k,i},k=1,\dots,N_p$, on the interval $[p^{\min}_i, p^{\max}_i]$ as
\begin{align}
p_{k,i} = \frac{p_i^{\max}+p_i^{\min}}{2} 
          + \frac{p_i^{\max}-p_i^{\min}}{2}\,L_k. 
          \label{eq:affine}
\end{align}
Then, the tensor-product Chebyshev--Lobatto grid on $\mathcal{X}$ is formed by
\begin{align}
\vec{p}_{\vec{k}} 
\;=\;
\bigl(\sigma^{(1)}_{\mathrm{gr}, k_1},\,S^{0, (1)}_{\mathrm{gr},k_2},\,\dots,\,\sigma^{(d)}_{\mathrm{gr}, k_{2d-1}}, S^{0, (d)}_{\mathrm{gr},k_{2d}} \bigr),
\quad
k_i \,\in\,\{1,\dots, N_{p} \}.
\end{align}
The grid has $N^{2d}_p$ nodes in total.

The numerical differentiation matrix \(D\in\mathbb{R}^{N_p\times N_p}\)
appearing in Eq.~(5.7) is obtained by evaluating the derivatives of
the Lagrange basis polynomials at the Chebyshev–Lobatto nodes.
According to \cite{doi:10.1137/1.9780898719598}, its entries are
\begin{align}
D_{11}
      &= \frac{2\bigl(N_p-1\bigr)^{2}+1}{6},
      && \label{eq:D00}\\[6pt]
D_{N_p,N_p}
      &= -\frac{2\bigl(N_p-1\bigr)^{2}+1}{6},
      && \label{eq:DNN}\\[6pt]
D_{kk}
      &= -\frac{L_k}{2\bigl(1-L_k^{2}\bigr)},
      &\qquad k &= 2,\dots,N_p-1,
      && \label{eq:Dkk}\\[6pt]
D_{ij}
      &=
      \displaystyle
      \dfrac{c_i}{c_j}\,
      \dfrac{(-1)^{i+j}}{L_i - L_j},
      & i \neq j,
      &\qquad i,j = 1,\dots,N_p,
      && \label{eq:Dij}
\end{align}
where the weights are
\begin{align}
c_i &=
\begin{cases}
2, & i = 1 \text{ or } i = N_p, \\[4pt]
1, & \text{otherwise}
\end{cases}
.
\label{eq:ci}
\end{align}
For an arbitrary interval $[p^{\min},p^{\max}]$, the scaled differentiation matrix is 
\begin{equation}
\widehat{D} = \frac{2}{p^{\max}-p^{\min}} D.
\end{equation}
The numerical differentiation error here is a truncation error arising from
discarding the higher-order modes of the Chebyshev expansion.

\subsection{Correlation matrix}\label{sec:corr_mat}
We consider three types of correlation matrices, as shown by our numerical demonstration later,  
significantly influence the bond dimensions of the TT representation for the characteristic function.
The first one is the constant correlation $\rho^{(\mathrm{const})}$ with all the off-diagonal elements equal to $0.5$.
The second one, the noisy correlation matrix $\rho^{(\mathrm{noise})}$, has off-diagonal elements randomly chosen from a uniform distribution between $0.4$ and $0.6$.
This case corresponds to adding a small amount of noise to $\rho^{(\mathrm{fixed})}$.
The third one, the random correlation $\rho^{(\mathrm{rand})}$, has off-diagonal elements randomly chosen from a uniform distribution between $0$ and $1$. 
Concretely, we fix these matrices as below:
\begin{equation}
\label{eq:matrices}
\resizebox{\textwidth}{!}{%
$
\begin{aligned}
\rho^{(\mathrm{const})} &=
\begin{pmatrix}
1   & 0.5 & 0.5 & 0.5 & 0.5 \\
0.5 & 1   & 0.5 & 0.5 & 0.5 \\
0.5 & 0.5 & 1   & 0.5 & 0.5 \\
0.5 & 0.5 & 0.5 & 1   & 0.5 \\
0.5 & 0.5 & 0.5 & 0.5 & 1
\end{pmatrix},
\quad
\rho^{(\mathrm{noise})} =
\begin{pmatrix}
1.0   & 0.472 & 0.595 & 0.453 & 0.554 \\
0.472 & 1.0   & 0.426 & 0.539 & 0.533 \\
0.595 & 0.426 & 1.0   & 0.531 & 0.462 \\
0.453 & 0.539 & 0.531 & 1.0   & 0.593 \\
0.554 & 0.533 & 0.462 & 0.593 & 1.0
\end{pmatrix},\\[1ex]
\quad
\rho^{(\mathrm{rand})} &=
\begin{pmatrix}
1.0   & 0.719 & 0.728 & 0.505 & 0.303 \\
0.719 & 1.0   & 0.394 & 0.132 & 0.515 \\
0.728 & 0.394 & 1.0   & 0.722 & 0.178 \\
0.505 & 0.132 & 0.722 & 1.0   & 0.401 \\
0.303 & 0.515 & 0.178 & 0.401 & 1.0
\end{pmatrix}
\end{aligned}
$}%
\end{equation}
In the random correlation case, after sampling the off-diagonal elements of the correlation matrix, we reorder the assets so that strongly correlated pairs are placed next to each other in the TT representation as far as possible. 
In the numerical demonstration, we have found that this reordering reduces the bond dimensions of TT for the characteristic function and improves the computational time of TCI.

\subsection{Settings of parameters}

We focus on the dependency of the option price and Greeks on $\sigma$ and $S_0$.
The ranges of these parameters are fixed as Table~\ref {tab:parameter_ranges}.
For each volatility parameter $\sigma_m$, we set the range to $\sigma_m \in [0.15, 0.25]$, centered at $0.2$, the typical value of Nikkei Stock Average Volatility Index \cite{nikkei_vi}, with a width of $\pm 0.05$, covering typical daily fluctuations. 
Only in the case of random correlations, we narrow the volatility parameter range to maintain accuracy.
For each initial asset price $S_{m,0}$, we set the range to $S_{m,0} \in [90, 120]$, representing a $\pm 20\%$ variation around $100$. The lower bound is set to $90$, rather than $80$, because the option price in our example becomes negligibly small for $S_0 < 90$.

\begin{table}[htb]
\centering
\setlength{\tabcolsep}{0.15pt}  
\footnotesize  
\begin{tabular}{cc}
\toprule
Parameter & Range \\
\midrule
 $\sigma$ & 
$\displaystyle \left\{
    \begin{array}{ll} 
    [0.175,\,0.225] & \text{(rand)} \\[1mm]
    [0.15,\,0.25] & \text{otherwise} 
    \end{array}
\right.$ \\
 $S_0$ & $[90,\,120]$ \\
\bottomrule
\end{tabular}
\caption{Parameter ranges for $\sigma$ and $S_0$.}
\label{tab:parameter_ranges}
\end{table}

\begin{table}[htb]
\centering
\scriptsize
\setlength{\tabcolsep}{2.4pt}
\begin{tabular}{ccccccccccccc}
\toprule
$d$ & $T$ & $r$ & $K$ & $\alpha$ & $\epsilon_{\mathrm{TCI}}$ &
\multicolumn{1}{c}{\begin{tabular}{c}
$\epsilon_{\mathrm{SVD}}(\tilde{\varphi})$ \\
$\epsilon_{\mathrm{SVD}}(\tilde{v})$
\end{tabular}} &
$\epsilon_{\mathrm{SVD}}\!\bigl(\tilde V_{(j_i,k_i,l_i)_{i=1}^d}\bigr)$ &
\multicolumn{1}{c}{
\begin{tabular}{c}
$\epsilon_{\mathrm{SVD}}\!\bigl(\tilde V_{(k_i,l_i)_{i=1}^d}\bigr)$ \\
$\epsilon_{\mathrm{SVD}}\!\bigl(\tilde{\nu},\tilde{\Delta},\tilde{\gamma}\bigr)$
\end{tabular}} &
$R_z$ &
$N_{z}$ & $N_{p}$ & $h$ \\
\midrule
5 & 1 & 0.01 & 100 & 
$\frac{5}{d}$ & 
$10^{-6}$ &
$\begin{aligned}
10^{-10} &: \mathrm{ND} \\
10^{-8} &: \mathrm{AN}
\end{aligned}$ &
$10^{-10}$ &
$\begin{aligned}
10^{-10} &: \mathrm{ND} \\
10^{-10} &: \mathrm{AN}
\end{aligned}$ &
25 &
127 & 100 &
$\begin{aligned}
0.001 &:  \nu \\
0.3   &: \Delta, \gamma
\end{aligned}$\\
\bottomrule
\end{tabular}
\caption{
Parameter settings used in TT-based option pricing, excluding the parameters $\vec{\sigma}$ and $\vec{S}_0$. The listed values include input parameters, TCI and SVD tolerance, finite integration ranges, grid discretization points, and finite-difference step sizes for fixed-seed-MC employed in numerical computations.}
\label{tab:other_parameter}
\end{table}

The other parameters are set as shown in Table~\ref{tab:other_parameter}.
We carefully select numerical tolerances to ensure accurate option pricing and Greeks calculation. 
The TCI tolerance is set to $\epsilon_{\mathrm{TCI}} = 10^{-6}$. 
For SVD, the tolerance $\epsilon_{\mathrm{SVD}}(\tilde{\varphi})$ and $\epsilon_{\mathrm{SVD}}(\tilde{v})$ are set to $10^{-10}$ for the ND approach and $10^{-8}$ for the AN approach.
In addition, the SVD tolerance for $\tilde{V}_{(j_i,k_i,l_i)_{i=1}^d}$ in the AN approach is set to $10^{-10}$. Likewise, we set the tolerance for $\tilde{V}_{(k_i,l_i)_{i=1}^d}$ to $10^{-10}$ in the ND approach. For the AN approach, the tolerances for TTs of Greeks, specifically $\tilde{\nu}_{(k_i,l_i)_{i=1}^d}$, $\tilde{\Delta}_{(k_i,l_i)_{i=1}^d}$, and $\tilde{\gamma}_{(k_i,l_i)_{i=1}^d}$, are also set to $10^{-10}$.
These parameter settings balance computational efficiency with accuracy. Specifically, the chosen tolerances keep computational costs low while ensuring the corresponding errors remain below the MC error.
We numerically verified that increasing the TCI tolerance from $10^{-6}$ to $10^{-5}$ significantly reduces the accuracy of the computed Greeks, underscoring the importance of setting an appropriate tolerance level. 

When dealing with random correlation matrices, we set $\epsilon_{\mathrm{SVD}}(\tilde{\varphi})$ and $\epsilon_{\mathrm{SVD}}(\tilde{v})$ to $10^{-8}$ because a tighter tolerance of $10^{-10}$ would lead to excessive memory usage during the SVD step in Eq.~\eqref{eq:elemntwise_phi_v}. 
This limitation arises from the computational complexity of the SVD, which scales as $O((\chi \chi')^3) \sim O(\chi^6)$, assuming the two TTs involved in the element-wise multiplication have maximum bond dimensions $\chi$ and $\chi'$, with $\chi' \simeq \chi$.
If SVD is performed using a sparse-matrix solver that avoids explicitly constructing the full matrix, the computational cost can be reduced to $O(\chi^5)$. However, this approach computes only a subset of singular values, making it impossible to directly estimate the truncation error.

The values for $\alpha$ were adopted from previous studies~\cite{kastoryano2022highlyefficienttensornetwork}. 
We have confirmed the stability of solutions under slight variations of $\alpha$. Lastly, the parameters $N_z$, $\vec{z}^{\min}$ and $\vec{z}^{\max}$ are selected to be sufficiently large, ensuring that the accuracy of our method remains comparable to that of the MC.

\subsection{Error evaluation}
Because no closed-form expressions are available for either the price or the Greeks of the min-call option, we treat results obtained from high-precision MC simulations as the ground truth. Specifically, we compute the option price and its Greeks using MV-based computations evaluated with $10^{8}$ MC paths. The accuracy of the Greeks produced by our proposed method is assessed using the root-mean-square error (RMSE) relative to these reference values.

Aside from this ground truth, we run the MC-based calculation of the price and Greeks with $n_{\mathrm{MC}} = 10^{6}$ simulation paths, a typical value in practice, and let the results be the benchmarks compared to our TT-based method.
We use the fixed-seed MC for Delta and Vega, as it produced smaller errors (i.e., results closer to the ground truth) than MV. 
On the other hand, we employ MV for Gamma due to its superior accuracy for estimating second-order Greeks.
In the following, we collectively refer to these benchmarking approaches as ``MC'' for simplicity. 
The error of MC are calculated as the standard error, namely $ \sigma/\sqrt{n_{\rm MC}}$ where $\sigma $ is the sample standard deviation of the discounted payoff in the respective runs.
We assess the accuracy of our method by checking whether its error is lower than that of the MC.

One issue is that the number of possible combinations of the parameter $\vec{\sigma}, \vec{S}^{0}$ is $N^{2d}_p = 100^{2d}$ where $N_p = 100$, and thus we cannot test all of them.
Instead, we randomly select 100 parameter combinations and compute the Greeks for each. 
We then compare the RMSE of our method against the errors of the MC under the same parameter settings.

\subsection{Software and hardware}
\texttt{TensorCrossInterpolation.jl}~\cite{N_ez_Fern_ndez_2025} was used for TT learning. 
The MC simulations were performed using~\texttt{tf-quant-finance}~\cite{monte}. 
Neither parallelization nor GPUs were used in any of the computations. 
All calculations were carried out on a CPU with about 1 TB of memory and a memory bandwidth of approximately 410 GB/s.

\section{Results of the numerical experiments}
\label{sec:experiments} 
We investigated three numerical approaches: the TT-based method with numerical differentiation (ND), the TT-based method with analytical differentiation (AN), and the MC-based method.
For each method, we present the computational complexity, computational time, and accuracy of option pricing and Greeks computations while the two parameters \(\vec{\sigma}\) and \(\vec{S_0}\) are varied in the five-asset case.
We show results for Greeks with respect to the first parameter $k_1$ or $l_1$. 
Table~\ref{table:table_res} summarizes these results.

The computational complexity and time are denoted by $c_{\mathrm{TT, ND}}$ and $t_{\mathrm{TT, ND}}$ for the ND approach, while they are represented by $c_{\mathrm{TT, AN}}$ and $t_{\mathrm{TT, AN}}$ for the AN approach. 
For 100 randomly selected parameter sets, the RMSE in TT-based option pricing and Greeks calculations is denoted by $e_{\mathrm{TT, ND}}$ for the ND approach and $e_{\mathrm{TT, AN}}$ for the AN approach.
For the same parameter sets, the standard error for MC-based option pricing or Greeks using $10^6$ paths is denoted by $e_{\mathrm{MC, 10^6}}$.
Additionally, the computational complexity and time of MC-based pricing and Greeks computations using $10^6$ paths are denoted by $c_{\mathrm{MC, 10^6}}$ and $t_{\mathrm{MC}}$. 
These computational times are the averages over 100 randomly selected parameter sets.

In these experiments, we utilize common TT representations for the characteristic function and payoff across methods, meaning that the TT representations of the option price are identical between ND and AN approaches.

\begin{figure*}[htbp]
    \centering
\includegraphics[width=1.0\linewidth]{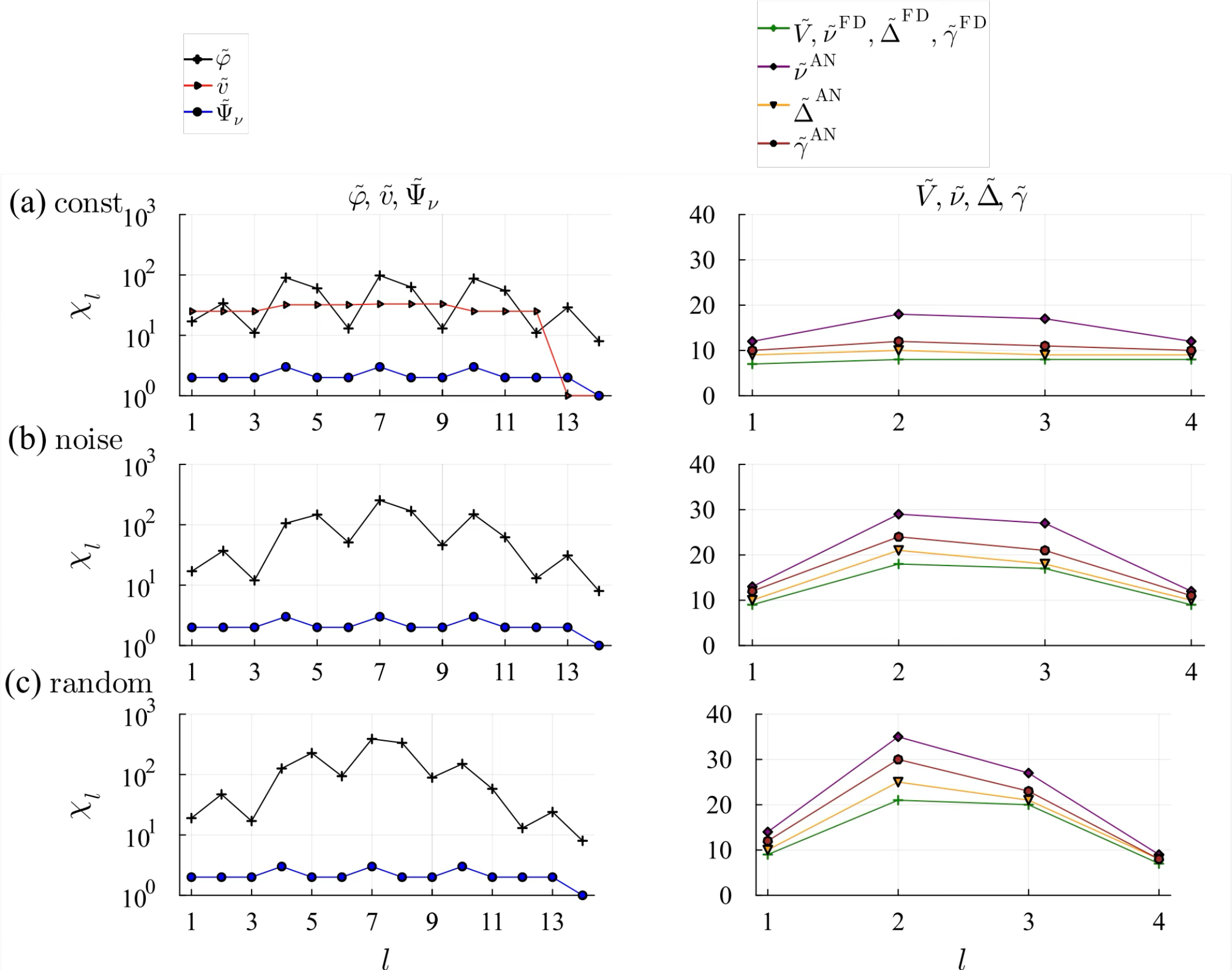}
    \caption{ 
    The bond dimensions are plotted against the bond index $l$ under three different correlation-matrix conditions: (a) constant, (b) noisy, and (c) random.
The left panels show the bond dimensions $\chi_l$ of the TTs of the functions $\varphi$, $\tilde{v}$, and $\Psi_{\nu}$, learned by TCI. 
For $\varphi$ and $\tilde{v}$, the TTs were truncated via SVD, whereas no SVD truncation was applied to $\Psi_{\nu}$. 
The bond dimensions of $\tilde{v}$ are shown only for the constant correlation case, as they are identical across all other cases. 
The right panels display the bond dimensions for the option price  ${V}$, Vega $\nu$,  Delta $\Delta$, and Gamma $\gamma$ as functions of bond index $l$. 
In each case, the bond dimensions of TT obtained via the ND approach are compared with those obtained from the AN approach.
       }
\label{fig:bonddim}
\end{figure*}

\subsection{Bond dimensions under different correlation matrices}
Figure \ref{fig:bonddim} (left) plots the bond dimensions in the TT representations after TCI and SVD for the characteristic function $\tilde{\varphi}$ and payoff function $\tilde{v}$.
For comparison, it also shows the bond dimensions of $\tilde{\Psi}_{\nu}$, which is computed via TCI (i.e., without an additional SVD step).
These bond dimensions largely determine the computational cost and numerical accuracy of the TT-based Fourier pricing and Greeks calculations.

The bond dimensions of $\tilde{\varphi}$ are highly sensitive to the correlation matrix.
Specifically, the maximal rank rises from roughly 100 for the constant case to about 150 for the noisy case and to nearly 400 for the random case. 
This growth reflects the increasing difficulty of compressing complicated inter-asset correlations within a one-dimensional chain of TT cores.

In contrast, the bond dimensions of the payoff tensor $\tilde{v}$ do not vary with the correlation structure, since it does not depend on the correlation matrix, and take relatively small values in all cases. 
The bond dimensions of $\tilde{\Psi}_{\nu}$ are significantly lower, with a maximum bond dimension of only 3, and this value remains almost unchanged regardless of the correlation matrix used.

Figure \ref{fig:bonddim} (right) presents the bond dimensions for TT representations of option prices and Greeks. 
Under the ND approach, these bond dimensions increase from around 10 with the constant correlation matrix to about 20 with the random one. 
In the AN approach, the bond dimensions for the option prices match those in the ND approach. 
However, for the Greeks, elementwise multiplication between $\tilde{V}_{(j_i,k_i,l_i)_{i=1}^{d}}$ and ($\Psi_{\nu}$, $\Psi_{\Delta}$, $\Psi_{\gamma}$) increases the complexity of the functions, thereby resulting in higher TT bond dimensions compared to the option prices.
Among all Greeks, Vega consistently exhibits the largest bond dimension, typically ranging from roughly 20 to 35 as the correlation structure becomes more complex.

\begin{table}[H]
\centering
\scalebox{0.68}{
\begin{tabular}{|c|ccc|ccc|ccc|}
\multicolumn{10}{c}{\textbf{(a) Constant correlation}} \\
\hline
 & $e_{\mathrm{TT,ND}}$ & $e_{\mathrm{TT,AN}}$ & $e_{\mathrm{MC},10^6}$ & $c_{\mathrm{TT,ND}}$ & $c_{\mathrm{TT,AN}}$ & $c_{\mathrm{MC},10^6}$ & $t_{\mathrm{TT,ND}}$ & $t_{\mathrm{TT,AN}}$ & $t_{\mathrm{MC},10^6}$ \\
\hline
$V$      & \multicolumn{2}{c|}{$5.61\times10^{-4}$} & $4.62\times10^{-3}$ & \multicolumn{2}{c|}{$199$} & \multirow{4}{*}{$ 5.0 \times 10^6$} & \multicolumn{2}{c|}{$3.78\times10^{-6}$} & $1.09\times10^{-1}$ \\
$\nu$    & $8.82\times10^{-3}$ & $9.32\times10^{-3}$ & $1.10\times10^{-2}$ &  & $750$ &  & 
& $6.97\times10^{-6}$ & $3.50\times10^{-1}$ \\
$\Delta$ & $3.65\times10^{-5}$ & $4.27\times10^{-5}$ & $1.27\times10^{-4}$ &  199 & $279$ &  & 
$3.78\times10^{-6}$
& $3.17\times10^{-6}$ & $2.62\times10^{-1}$ \\
$\gamma$ & $5.48\times10^{-6}$ & $6.41\times10^{-6}$ & $2.56\times10^{-5}$ &  & $382$ &  & 
& $8.40\times10^{-6}$ & $9.10\times10^{-1}$ \\
\hline
\end{tabular}
}

\vspace{1.5em}

\scalebox{0.68}{
\begin{tabular}{|c|ccc|ccc|ccc|}
\multicolumn{10}{c}{\textbf{(b) Constant correlation with added noise}} \\
\hline
 & $e_{\mathrm{TT,ND}}$ & $e_{\mathrm{TT,AN}}$ & $e_{\mathrm{MC},10^6}$ & $c_{\mathrm{TT,ND}}$ & $c_{\mathrm{TT,AN}}$ & $c_{\mathrm{MC},10^6}$ & $t_{\mathrm{TT,ND}}$ & $t_{\mathrm{TT,AN}}$ & $t_{\mathrm{MC},10^6}$ \\
\hline
$V$      & \multicolumn{2}{c|}{$7.81\times10^{-4}$} & $4.87\times10^{-3}$ & \multicolumn{2}{c|}{$639$} & \multirow{4}{*}{$ 5.0 \times 10^6$} & \multicolumn{2}{c|}{$3.78\times10^{-6}$} & $1.11\times10^{-1}$ \\
$\nu$    & $9.32\times10^{-3}$ & $3.47\times10^{-2}$ & $1.11\times10^{-2}$ &  & $1509$ &  & 
& $9.85\times10^{-6}$ & $2.45\times10^{-1}$ \\
$\Delta$ & $5.04\times10^{-5}$ & $5.41\times10^{-5}$ & $1.62\times10^{-4}$ &  639& $788$  &  &$3.78\times10^{-6}$ 
& $4.71\times10^{-6}$ & $3.52\times10^{-1}$ \\
$\gamma$ & $8.59\times10^{-6}$ & $9.08\times10^{-6}$ & $2.98\times10^{-5}$ &  & $1046$ &  & 
& $3.98\times10^{-6}$ & $9.98\times10^{-1}$ \\
\hline
\end{tabular}
}

\vspace{1.5em}

\scalebox{0.68}{
\begin{tabular}{|c|ccc|ccc|ccc|}
\multicolumn{10}{c}{\textbf{(c) Random correlation}} \\
\hline
 & $e_{\mathrm{TT,ND}}$ & $e_{\mathrm{TT,AN}}$ & $e_{\mathrm{MC},10^6}$ & $c_{\mathrm{TT,ND}}$ & $c_{\mathrm{TT,AN}}$ & $c_{\mathrm{MC},10^6}$ & $t_{\mathrm{TT,ND}}$ & $t_{\mathrm{TT,AN}}$ & $t_{\mathrm{MC},10^6}$ \\
\hline
$V$      & \multicolumn{2}{c|}{$7.89\times10^{-4}$} & $4.61\times10^{-3}$ & \multicolumn{2}{c|}{$765$} & \multirow{4}{*}{$ 5.0 \times 10^6$} & \multicolumn{2}{c|}{$5.93\times10^{-6}$} & $1.05\times10^{-1}$ \\
$\nu$    & $1.78\times10^{-2}$ & $4.29\times10^{-2}$ & $8.28\times10^{-3}$ &  & $1701$ &  & 
& $1.17\times10^{-5}$ & $2.71\times10^{-1}$ \\
$\Delta$ & $1.31\times10^{-4}$ & $1.34\times10^{-4}$ & $1.27\times10^{-4}$ &  765& $961$  &  & $5.93\times10^{-6}$ 
& $9.47\times10^{-6}$ & $4.01\times10^{-1}$ \\
$\gamma$ & $2.59\times10^{-5}$ & $2.57\times10^{-5}$ & $6.92\times10^{-5}$ &  & $1254$ &  & 
& $8.41\times10^{-6}$ & $1.05$ \\
\hline
\end{tabular}
}
\caption{
Numerical results comparing errors $(e)$, computational complexities $(c)$, and computation times $(t)$ of TT-based ND and AN approaches versus the MC benchmark with $10^6$ paths.
Results are shown for three correlation structures: (a) constant, (b) noisy, and (c) random. 
Note that the ND and AN approaches share the same accuracy for the option price $V$, as well as the same computational complexities and times for $V$. 
Additionally, for the ND approach, computational complexity and time remain the same across all Greeks and option prices within each correlation case.
}
\label{table:table_res}
\end{table}

\subsection{Accuracy, computational complexity, and time}
Tables~\ref{table:table_res}(a)–(c) summarize the numerical results of our experiments, demonstrating that both TT-based methods, ND and AN, significantly outperform the MC in terms of computational complexity and time when considering only the online phase.
Specifically, these TT-based methods reduce computational complexity by approximately $10^{3}-10^{4}$ times and achieve execution-time speed-ups on the order of $10^{4}-10^{5}$ compared to MC. 
Note that the improvement factor in computational complexity differs from that in runtime by one order of magnitude, since the complexity of MC is measured as $d N_{\mathrm{path}}$, whereas for the TT-based methods, we count the actual arithmetic operations involved.
Even with such a counting method potentially disadvantageous for TT-based methods, TT-based methods still exhibit a computational advantage over the MC.
No notable runtime differences were observed between the ND and AN approaches.

Both TT-based methods generally maintain accuracy comparable to MC. 
However, accuracy declines in the random correlation case, especially for Vega. 
This highlights two distinct factors influencing the accuracy of TT representations. 
First, the random correlation case generally yields larger errors 
compared to the other two cases. 
This occurs because random correlation structures require significantly higher bond dimensions to accurately represent the characteristic function, making TCI more susceptible to becoming trapped in local minima. 
Second, Vega appears inherently more challenging for accurate TT approximation than Delta and Gamma, regardless of the correlation structure, as shown in Figure~\ref{fig:bonddim} (right). 



\begin{figure}[h]
    \centering
    \includegraphics[width=1.0\linewidth]{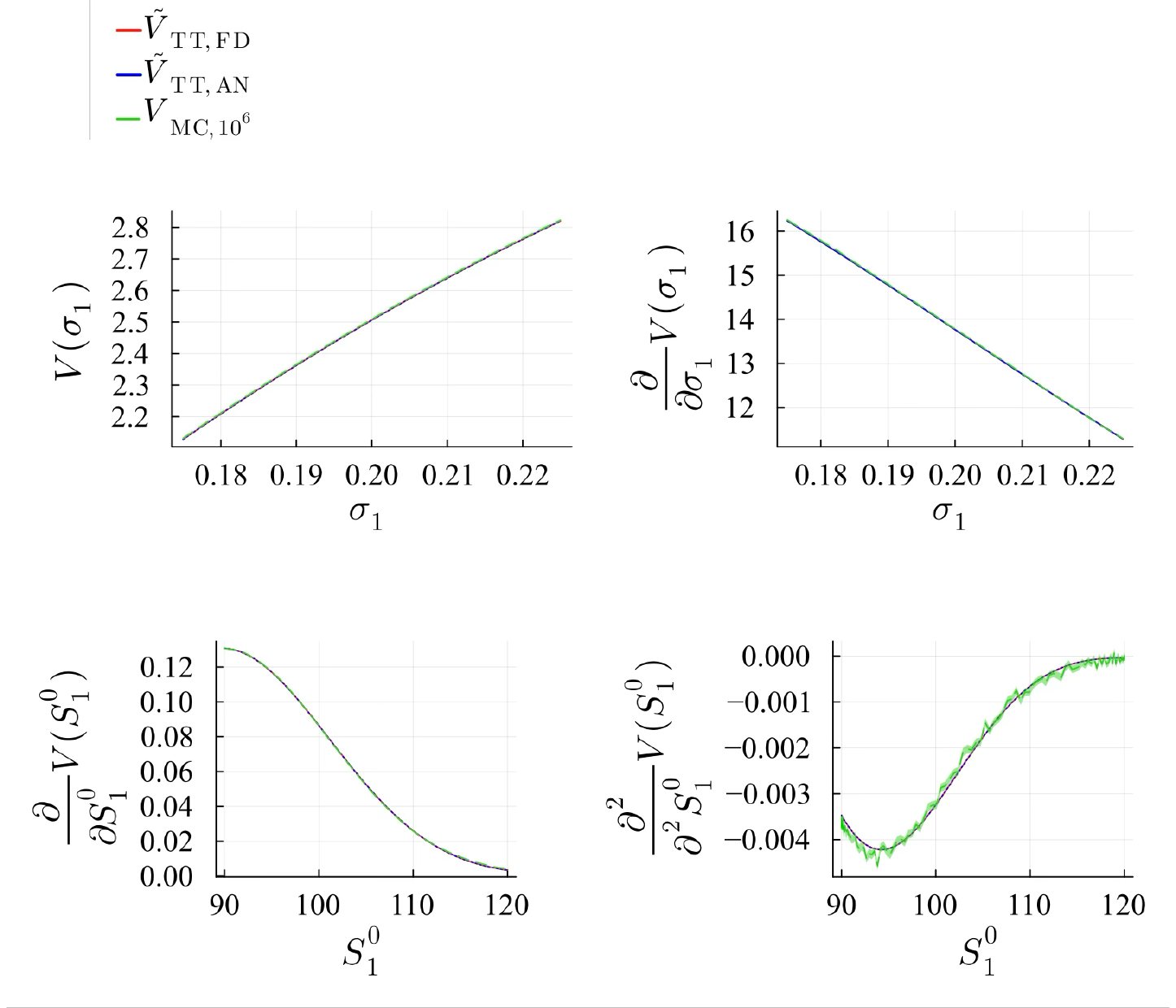}
    \caption{
    One-dimensional plots of the option price $V$ (top left), Vega $\nu$ (top right), Delta $\Delta$ (bottom left), and Gamma $\gamma$ (bottom right) as functions of either $\sigma_1$ or $S^0_1$ in the random correlation case.
   In each plot, $\sigma_2,S^0_2,\dots,\sigma_5,S^0_5$ and all other input parameters are fixed at the values listed in Table~\ref{tab:other_parameter}.
    For the price $V$, the same TT representation is used for both the ND and AN approaches.
    }
    \label{fig:greeks_1d}
\end{figure}

While the error indicators summarized in Table~\ref{table:table_res} are worse for the TT-based methods than the MC-based method in some cases, basically, these methods yield very close results. 
Figure~\ref{fig:greeks_1d} presents representative one-dimensional slices in the random correlation case. We plot the option price and Vega along the $\sigma_1$-axis and Delta and Gamma along the $S^{0}_1$-axis, fixing other $\sigma_m$ and $S^0_m$ to some values. 
Overall, both the AN and ND approaches show excellent agreement with MC across all the plots, including Vega.
We have found the following tendency: although Vega in the TT-based methods tends to have a slightly lower accuracy than MC in the region where the value of Vega is relatively small, it matches the ground truth closely in the region where the Vega value is large relative to the error.
Indeed, it is practically crucial to be accurate in the region of large Vega.

In a direct comparison between the ND and AN approaches, we find that the ND approach generally achieves higher overall accuracy. 
While the AN approach eliminates numerical differentiation errors, caused by truncating Chebyshev nodes, it still requires constructing separate TT representations for each Greek (Delta, Vega, Gamma). Forming these representations entails elementwise  multiplications of tensors involving the additional tensors $\tilde{\Psi}_\nu$, $\tilde{\Psi}_\Delta$, and $\tilde{\Psi}_\gamma$, 
and the accompanying compression operations (e.g., TCI and SVD) inevitably introduce additional TT approximation errors.
These additional approximation errors ultimately outweigh the benefits gained by avoiding numerical differentiation.

\section{Discussion}\label{sec:discussion}
We proposed the two TT-based methods for efficiently computing option Greeks within FT-based pricing for multi-asset options under the BS model. 
The first approach involves directly applying numerical differential operators within the TT cores representing the option price. 
The second approach analytically differentiates the Fourier pricing formula and subsequently compresses the resulting analytical formulas into the TT representation.

To evaluate these approaches, we test them on a five-asset min-call option under several correlation matrices, the constant, noisy, and random ones, in the characteristic function. 
Both TT-based methods demonstrate a substantial advantage in terms of computational complexity and runtime compared to MC with $10^{6}$ paths, considering only the online phase. 
Furthermore, the accuracy of these approaches is generally comparable to or better than these MC benchmarks.

A minor exception appears for Vega under randomly varying correlations, where the TT-based methods show slightly larger errors than MC. 
The discrepancies become noticeable only in regions where Vega is small.
In contrast, in the region with large values of Vega and thus of practical importance, the TT-based methods show better accuracy.

The ND approach is simpler in implementation and often more accurate than the AN approach. 
ND requires constructing only a single TT representation of the option price. AN eliminates numerical-differentiation errors but requires building a separate TT for each of the Greeks and performing extra offline preprocessing. Consequently, in terms of offline computational cost, 
ND is generally preferable.

Several promising directions remain open for future research. 
One important extension is moving beyond the BS model to more general models, such as L\'{e}vy models~\cite{RePEc:spr:finsto:v:2:y:1997:i:1:p:41-68, RePEc:ucp:jnlbus:v:63:y:1990:i:4:p:511-24}. 
Additionally, exploring the TT-based method for more computationally intensive instruments, such as American or Bermudan options, is also important. 
Although a previous study has proposed a tensor completion method for these types of options~\cite{doi:10.1137/19M1244172}, it remains unclear whether our proposed method in this study can efficiently handle them.
Moreover, although the current study has focused primarily on volatility $\vec{\sigma}$ and initial prices $\vec{S}_{0}$ as parameters that tend to vary most significantly during market hours, incorporating additional parameters such as maturity $T$ and interest rate $r$ into TTs is worth considering.
It would be needed to investigate whether the beneficial low-bond dimension structure persists as the number of parameters increases.

Finally, our current implementation evaluates prices and Greeks only on a fixed parameter grid (i.e., Chebyshve nodes). 
A previous study demonstrates that combining TT representations with interpolation techniques, such as Chebyshev interpolation, enables continuous evaluations at arbitrary parameter points~\cite{doi:10.1137/19M1244172}. 
Integrating such an interpolation polynomial into the proposed TT-based approach represents a promising direction.

\section*{Acknowledgments}
This work was supported by the Center of Innovation for Sustainable Quantum AI (JST Grant Number JPMJPF2221).
T.O. acknowledges the support from JSPS KAKENHI (Grant Nos. 23H03818 and 22K18682), the Endowed Project for Quantum Software Research and Education, The University of Tokyo (https://qsw.phys.s.u-tokyo.ac.jp/).
K.M. is supported by MEXT Quantum Leap Flagship Program (MEXT Q-LEAP) Grant no. JPMXS0120319794 and JST COI-NEXT Program Grant No. JPMJPF2014.
We used ChatGPT (GPT-4.5, OpenAI) to assist in proofreading the manuscript, specifically for checking spelling, grammar, and improving the clarity and style of the text. 
No original content was generated by the AI tool.

\bibliographystyle{siamplain}
\bibliography{references}
\end{document}

%% file: ex_shared.tex
\usepackage{lipsum}
\usepackage{amsfonts}
\usepackage{graphicx}
\usepackage{epstopdf}
\usepackage{algorithmic}
\ifpdf
  \DeclareGraphicsExtensions{.eps,.pdf,.png,.jpg}
\else
  \DeclareGraphicsExtensions{.eps}
\fi

\newsiamremark{remark}{Remark}
\newsiamremark{hypothesis}{Hypothesis}
\crefname{hypothesis}{Hypothesis}{Hypotheses}
\newsiamthm{claim}{Claim}

\headers{Greeks with tensor trains}{R. Sakurai, K. Miyamoto, and T. Okubo}

\title{Tensor train representations of Greeks for Fourier-based option pricing of multi-asset options \thanks{Submitted to the editors DATE.
\funding{This work is supported by the Center of Innovation for Sustainable Quantum AI (JST Grant Number JPMJPF2221). 
KM is supported by MEXT Quantum Leap Flagship Program (MEXT Q-LEAP) Grant No. JPMXS0120319794 and JST COI-NEXT Program Grant No. JPMJPF2014.}}}

\author{Rihito Sakurai\thanks{The University of Tokyo 
  (\email{sakurairihito@gmail.com}).}
\and Koichi Miyamoto\thanks{Center for Quantum Information and Quantum Biology, The University of Osaka 
  (\email{miyamoto.kouichi.qiqb@osaka-u.oac.jp}).}
\and Tsuyoshi Okubo\footnotemark[2] \thanks{Institute for Physics of Intelligence, University of Tokyo (\email{t-okubo@phys.s.u-tokyo.ac.jp}).}
}
\usepackage{amsopn}
